\documentclass[10pt,letterpaper]{article}
\usepackage[top=0.85in,left=2.75in,footskip=0.75in]{geometry}

\usepackage{amsmath,amssymb}

\usepackage{changepage}

\usepackage[utf8]{inputenc}

\usepackage{textcomp,marvosym}

\usepackage{nameref}

\usepackage{microtype}
\DisableLigatures[f]{encoding = *, family = * }

\usepackage[table]{xcolor}
\usepackage{tabularx}

\usepackage{array}

\newcolumntype{+}{!{\vrule width 2pt}}

\newlength\savedwidth

\raggedright
\usepackage[aboveskip=1pt,labelfont=bf,labelsep=period,justification=raggedright,singlelinecheck=off]{caption}

\usepackage[citestyle=numeric,
            backend=bibtex,
            bibstyle=authoryear,
            sorting=none,
            sortcites,
            firstinits,
            terseinits,
            maxbibnames=6,
            dashed=false]{biblatex}
\bibliography{bibliography}
\usepackage[colorlinks]{hyperref}

\DeclareNameAlias{sortname}{last-first}

\DeclareFieldFormat{title}{#1}
\DeclareFieldFormat[article,inbook,incollection,inproceedings,patent,
                    thesis,unpublished]{title}{#1}
\DeclareFieldFormat{journaltitle}{#1}
\DeclareFieldFormat{issuetitle}{#1}
\DeclareFieldFormat{maintitle}{#1}

\renewbibmacro*{in:}{%
  \ifentrytype{article}{}{\printtext{\bibstring{in}\intitlepunct}}}

\DeclareFieldFormat{pages}{#1}

\renewbibmacro*{volume+number+eid}{%
  \setunit*{\addspace}%
  \printfield{volume}}

\DeclareFieldFormat{labelnumberwidth}{#1\adddot}
\defbibenvironment{bibliography}
  {\list
     {\printtext[labelnumberwidth]{%
      \printfield{prefixnumber}%
      \printfield{labelnumber}}}
     {\setlength{\labelwidth}{\labelnumberwidth}%
      \setlength{\leftmargin}{\labelwidth}%
      \setlength{\labelsep}{\biblabelsep}%
      \addtolength{\leftmargin}{\labelsep}%
      \setlength{\itemsep}{\bibitemsep}%
      \setlength{\parsep}{\bibparsep}}%
      }
  {\endlist}
  {\item}

\AtEveryBibitem{\clearfield{extrayear}}

\usepackage{lastpage,fancyhdr,graphicx}
\usepackage{epstopdf}
\fancyheadoffset[L]{2.25in}
\fancyfootoffset[L]{2.25in}
\usepackage[capitalize, nameinlink]{cleveref}
\usepackage[inline]{enumitem}
\usepackage[binary-units, range-units=single, range-phrase=~to~, list-units=single]{siunitx}
\usepackage{subcaption}
\usepackage{amsthm}
\usepackage{thmtools}

\declaretheoremstyle[
notefont=\bfseries, notebraces={}{},
bodyfont=\normalfont,
headformat=\NAME~\NUMBER~---~\NOTE%
]{nopar}
\declaretheorem[style=nopar]{definition}

\usepackage[export]{adjustbox}
\usepackage{multirow}
\usepackage{booktabs}

\usepackage{pgf-umlsd}
\usepackage{tikz}
\usepackage{tikzpeople}
\usetikzlibrary{shapes,arrows.meta,positioning,automata,matrix,shapes.symbols,trees}
\usepackage{tikz-qtree}
\usepackage[]{todonotes}
\reversemarginpar  

\newenvironment{acks}{
  \section*{Acknowledgements}
}{}

\begin{document}
\vspace*{0.2in}

\begin{flushleft}
{\Large
\textbf\newline{Impact of delayed response on Wearable Cognitive Assistance} 
}
\newline
\\
Manuel~{Olguín~Muñoz}\textsuperscript{1, *},
Roberta~Klatzky\textsuperscript{2, 3},
Junjue~Wang\textsuperscript{4},
Padmanabhan~Pillai\textsuperscript{5},
Mahadev~Satyanarayanan\textsuperscript{4},
James~Gross\textsuperscript{1}
\\
\bigskip
\begin{tabularx}{\textwidth}{lX}
  \textbf{1} & Division of Information~Science and Engineering, School of Electrical~Engineering and Computer~Science, KTH Royal Institute~of~Technology, Stockholm, Sweden\\
  \textbf{2} & Department of Psychology, Carnegie~Mellon~University, Pittsburgh, PA, USA\\
  \textbf{3} & Human-Computer Interaction Institute, Carnegie~Mellon~University, Pittsburgh, PA, USA\\
  \textbf{4} & School of Computer Science, Carnegie~Mellon~University, Pittsburgh, PA, USA\\
  \textbf{5} & Intel Labs, Pittsburgh, PA, USA\\
\end{tabularx}
\bigskip

\begin{tabularx}{\textwidth}{lX}
  * & Corresponding author: \href{mailto:molguin@kth.se}{\texttt{molguin@kth.se}}\\
\end{tabularx}

\end{flushleft}

\section*{Abstract}

Wearable Cognitive Assistants (WCA) are anticipated to become a widely-used application class, in conjunction with emerging network infrastructures like 5G that incorporate edge computing capabilities.
While prototypical studies of such applications exist today, the relationship between infrastructure service provisioning and its implication for WCA usability is largely unexplored despite the relevance that these applications have for future networks.
This paper presents an experimental study assessing how WCA users react to varying end-to-end delays induced by the application pipeline or infrastructure.
Participants interacted directly with an instrumented task-guidance WCA as delays were introduced into the system in a controllable fashion.
System and task state were tracked in real time, and biometric data from wearable sensors on the participants were recorded.
Our results show that periods of extended system delay cause users to correspondingly (and substantially) slow down in their guided task execution, an effect that persists for a time after the system returns to a more responsive state.
Furthermore, the slow-down in task execution is correlated with a personality trait, neuroticism, associated with intolerance for time delays.
We show that our results implicate impaired cognitive planning, as contrasted with resource depletion or emotional arousal, as the reason for slowed user task executions under system delay.
The findings have several implications for the design and operation of WCA applications as well as computational and communication infrastructure, and additionally for the development of performance analysis tools for WCA.\@

\section{Introduction}

Wearable Cognitive Assistants, \emph{WCA} for short, have recently started to garner attention from the research community~\autocite{Ha:TowardsWearableCogAssist,Chen:EarlyImplementation}.
They represent a novel category of highly interactive and context-sensitive augmented reality applications, that aim to augment human cognition in both day-to-day tasks and professional settings.
Their mode of operation is analogous to how GPS navigation systems guide drivers, by seamlessly providing relevant instructions and feedback relating to the current task at hand.
Note that this implies seamless interaction with the context of the user --- at no moment does the user need to trigger an update explicitly, as the application is constantly tracking the state of the target task.
An example is an \textcite{IKEAAssistant} which monitors the assembly of a piece of furniture in real time, providing timely, step-by-step feedback to guide the user toward completion.

WCA systems show great potential for two main use-cases.
One is providing quality of life improvements to the millions of people around the world affected by some form of cognitive decline. 
WCA can, for instance, provide assistance to people recovering from traumatic brain injuries, smoothly guiding them through day-to-day interactions with the world which would otherwise be extremely challenging. 

The second use-case is as a companion tool for specialists and as a means for guiding their training. 
Non-wearable augmented reality and cognitive assistance systems have already been proven to be valuable tools in the industrial workplace~\autocite{funk2015caworkplace,gorecky2011cognito}.
Detethering this assistance from its current fixed location will surely make it available to many more fields. 

Based on these use cases, we identify three main requirements for WCA:\@
\begin{enumerate}
    \item\label{item:pervasive} WCA systems should be available whenever the user requires them, without being tethered to a particular physical location. Assistants need to be pervasive and mobile.

    \item\label{item:interaction} Interaction with the system should be immersive and seamless, i.e.\ the assistant should be able to analyze the current context and automatically provide relevant feedback without explicit commands from the user.
    In this sense, WCA is expected to operate much like a human assistant would, by observing the performance of the user and offering guidance proactively.

    \item\label{item:lowlatency} Feedback should be ``quick'', relative to the task at hand. 
    This requirement is further strengthened by the previously mentioned ``seamless interaction'' characteristic of these systems.
    This means that users will have expectations of constant, immediate feedback as they progress through the task.
    In the case of a step-by-step task like IKEA, delayed feedback might simply confuse or distract the user.
    However, in a highly interactive task like a \emph{Ping-Pong assistant}~\autocite{PingPongAssistant, Chen:EarlyImplementation} late guidance is at best a nuisance and at worst a severe handicap.
\end{enumerate}

\cref{item:pervasive} implies use of lightweight and low-power devices, preferably a wearable device that frees both hands for work.
\cref{item:interaction}, on the other hand, suggests a level of context sensitivity and proactivity that can only be provided by real-time analysis of sensor inputs such as video and audio feeds.
The compute-intensive processing suggested by \cref{item:interaction} cannot be met by the lightweight wearable devices suggested by \cref{item:pervasive}.
Only by offloading computation from a wearable device to cloud-based or edge-based infrastructure can this circle be squared.
However, offloading implies an extended end-to-end pipeline with many potential sources of queueing, transmission, and processing delays.
\cref{item:lowlatency} therefore emerges as a key concern, requiring deep understanding of the impact of end-to-end delays on WCA users.

\cref{item:lowlatency} forms the base motivation for the research presented in this paper.
We still have a very limited understanding of how humans react to delays in these systems --- specifically, how changes in system responsiveness impact users.
\emph{System responsiveness} here denotes a qualitative scale ranging from \emph{high} (that is, not subject to delay or subject to negligible delay with respect to human perception) to \emph{low} (i.e.\ subject to considerable delay).
Characterizing the relationships between system responsiveness and user behavior and experience is of paramount importance for the design and evaluation of these applications.
A clear understanding of these relationships would allow, for instance, for the development of strategies for load balancing and optimization for large-scale deployments of WCA.\@

This paper builds upon preliminary work in the field of time perception and delay characterization of WCA.\@
We expand upon the findings of \textcite{Ha:TowardsWearableCogAssist}, who identified the need for low-latency offloading in WCA, and of \textcite{Chen:AnEmpiricalStudyOfLatency}, who outlined the bounds for ``noticeable'' and ``unbearable'' latencies in these systems.
While these bounds present a general understanding of when it is likely that a user will drop an application, they do not provide any insights as to what happens \emph{before} that --- i.e. how human behavior changes with system responsiveness.
We aim to tackle this question through the characterization of human responses to delays in the application pipeline, using latencies in the range defined by the previously established bounds.
This is an important step toward a more systematic understanding of human behavior in this domain.

We present in this paper an experimental WCA test-bed of our design and making.
This test-bed was subsequently employed in a study in which undergraduate students were asked to interact with and follow the instructions given to them by a cognitive assistant.
Unbeknownst to the participants, we altered the responsiveness of the system in real-time and recorded the resulting behavioral and physiological reactions. 
The participants wore an array of biometric sensors measuring physiological responses that have proven useful in assessing cognitive workload during human-computer interaction~\autocite{haapalainen2010psycho,kumar2016measurement} such as heart rate and EEG.\@

Our results indicate that reduced responsiveness in WCA systems leads to a disruption of participants' cognitive plan for the task.
This is evidenced by an emergent pacing effect on user actions as system responsiveness is reduced.
While it would seem self-evident that users take longer to complete a task while using a system with low responsiveness --- as they have to wait longer for new instructions --- our study found that user slow-down represents a source of substantial additional delay.
To be more precise, the data indicate that users slow down not only because they have to wait for the system to catch up, but that their reactions to new instructions is also delayed.
Furthermore, this effect persists for a while after system response improves and is modulated by intrinsic personality traits, in particular, \emph{neuroticism}~\autocite{john1999:bfi}, which has previously been connected to intolerance for time delay~\autocite{hirsh2008delay}. 

We believe that these results provide concrete and relevant implications for WCA design, deployment, and optimization.
One example is the behavioral slow-down, as it extends application runtime significantly, and thus has clear and direct implications for resource and power consumption.
Another is the fact that the adverse effects of delay on users do not immediately subside as delay is diminished --- this has potential consequences for resource allocation strategies.
Moreover, in multi-user scenarios, the dependency of user slow-down effects on delay mean efficient resource allocation across applications potentially looks different from what could be considered ``fair''.

Our hope is that the results we provide might prove useful for the understanding and optimization of deployments of WCA.\@
These results represent unexpected, valuable findings, which can be employed to model and understand how users interact with latency in applications and systems, and develop resource allocation and power optimization strategies.
Additionally, we hope that the results we provide might pave the way for the improvement of performance evaluation tools such as our previous work in \textcite{olguin:2018, olguin:2019}.
Such systems would greatly benefit from this knowledge, as it would allow for the design and implementation of realistic models of human behavior, making highly accurate benchmarks a reality in the domain of WCA.\@

The structure of this paper is as follows.
We describe the existing body of research around time perception and the effects of delay on human performance in \cref{sec:background}.
\cref{sec:experimentaldesign} presents the experimental design, measures, and specific protocol.
Then, in \cref{sec:results} we detail the results of our experiment.
Implications for further modeling of the effects of delay are presented in \cref{sec:discussion} before finally concluding the paper in \cref{sec:conclusion}.

\section{Background and Related Work}\label{sec:background}
\subsection{Time perception in computing systems}

The question of how people respond to delay in a computer system is grounded in how people perceive time.
Time perception has been described as regulated by an attentional gate that, when opened, starts a cognitive pulse counter~\autocite{zakay1995attentional,zakay1996role}.
More recent research indicates, however, that duration perception is highly malleable and the result of multiple timing mechanisms found in overlapping, flexible neural systems~\autocite{bruno2016multiple,wiener2011multiple}. 
The estimation of an event's duration varies with context of various types
\begin{enumerate*}[label={(\roman*)}, before=\unskip{: }, itemjoin={{; }}, itemjoin*={{; and }}]
    \item events subsequent to a long or short interval are contracted or extended, respectively~\autocite{heron2012duration}
    \item repeated events tend to be perceived as shorter than novel ones~\autocite{matthews2011stimulus}
    \item arousal can expand durations~\autocite{droit2011emotion}.
\end{enumerate*}

Expectations play a critical role in time perception as well~\autocite{zakay1995attentional,zakay1996role}.
It has been shown that people have a general tendency to be hypersensitive to delays in worse-than-expected states, and under-sensitive to meeting or exceeding expectations~\autocite{Loewenstein1992anomaliesintertemporalchoice}.
Accordingly, failures to meet expected fast response times tend to be experienced as highly negative, whereas fast latencies are not noticed.
Violations of expectancy have a strong impact on the acceptability of computer systems.
Users of a computer system anticipate the latency for events, for which the standards only become more stringent as systems improve in response time.
In immersive systems like WCA, which aim to provide seamless interaction, delays are particularly noticeable.

It has long been recognized that slow system response times can undermine cognitive processing, slow the pace of users, and lead to stereotyped behavior and errors, as well as cause negative emotional consequences~\autocite{dabrowsky:2011:40years}.
However, standards for what constitute tolerable delays have changed dramatically compared to three decades ago, when delays on the order of \SI{10}{\second} were deemed acceptable~\autocite{nielsen1994usability, shneiderman2016designing, seow2008designing}.
Today's user context, and WCA in particular, often demand response times orders of magnitude shorter.

For WCA the acceptable range for latencies was explored by \textcite{Chen:AnEmpiricalStudyOfLatency}, by constructing assistants for tasks with a range of time constraints, including step-by-step tasks and more interactive contexts like playing Ping-Pong against a human opponent.
They then proposed a latency tolerance zone according to the task demands.
For an essentially self-paced task like LEGO assembly, they found two key ranges of latency; unnoticeable, \SIrange{0}{0.6}{\second}; and impaired, \SIrange{0.6}{2.7}{\second}. 
Beyond that, users could begin to show the negative outcomes previously catalogued~\autocite{dabrowsky:2011:40years}.

\subsection{Potential Mechanisms Relating Delay to Human Performance}\label{ssec:potentialmechs}

While behavioral changes and negative interaction outcomes have been well documented in prior research on system delay, the specific  mechanisms that mediate these outcomes are less well understood. 
These mechanisms could be cognitive or emotional in origin.

Research on cognitive and motor planning suggests that delay may move users from relatively automatic to more attention-demanding processing.
Cognitive and motor tasks are commonly described as a hierarchical system, progressing from high-level goals to the sequence of commands that accomplishes them.

As competency in a task increases, execution of the hierarchy becomes increasingly automated.
Automatization has been described from a computational perspective in {Anderson's ACT-R}~model as the compiling of multiple productions into one~\autocite{neves1981knowledge}.
Neural measurements indicate that with automaticity, control moves from frontal brain areas to more posterior ones~\autocite{jeon2015degree,puttemans2005changes}, and similar distinctions have been related to temporal processing~\autocite{lewis2003distinct,koch2009neural,lee2019limiting}.

Although activities guided by a WCA are not simple motor actions, immediate feedback after each of a series of repeated actions should promote development and automatic execution of a hierarchical plan.
Delays, in contrast, would disrupt such a plan through the loss of automated control~\autocite{lee2019limiting}.

An alternative view of delay effects appeals to emotional systems rather than cognitive processes.
As users of a system become emotionally aroused by delay, they may be subject to generalized arousal, causing decrements in performance~\autocite{lee2019limiting}.

A third potential explanation of delay effects is what has been called ``ego depletion'', the notion that expending effort on self-control eliminates resources needed for further effort~\autocite{baumeister74tice,lin2020strong}.

The various processing accounts of delay effects predict different outcomes, which we will consider in the context of the current data.
If delay increases attentional demands on cognitive processes, responses should be slowed and errors expected, particularly on time-critical tasks.
Generalized arousal triggered by emotional stress from delay should emerge in physiological measures, such as increased heart rate or skin conductivity.
Arousal can also reduce movement smoothness or add erratic gestures~\autocite{pijpers2003anxiety}.
Ego depletion has been found to produce premature responses culminating in error~\autocite{lin2020strong}, or to lead to abandoning a task entirely~\autocite{baumeister74tice}. 

Over-arching prescriptions for tolerable system response time have not tended to take into account individual differences in users with respect to salient variables like cognitive ability or personality. 
Relevant research can be found in studies of delay discounting, the tendency to devalue rewards for which one must wait.
High discounting rates, indicative of waiting intolerance, have been associated with negative social and academic outcomes.
\textcite{hirsh2008delay} found that higher discounting was associated with extraversion among those with low cognitive function, whereas lower discounting was associated with emotional stability (low neuroticism) for people with high cognitive function.
Among computer system users who tend to have relatively high cognitive ability (which presumably describes the present experimental population), this points to neuroticism as a personality factor that might modulate tolerance for waiting. Extraversion could also  be  a moderating factor among the broader target audience of WCA, which are intended for relatively inexperienced users of an application.
These and other measures of individual variation were considered here.

\section{Experimental Design}\label{sec:experimentaldesign}

\begin{figure}[h]
  \centering
  \includegraphics[width=.8\textwidth]{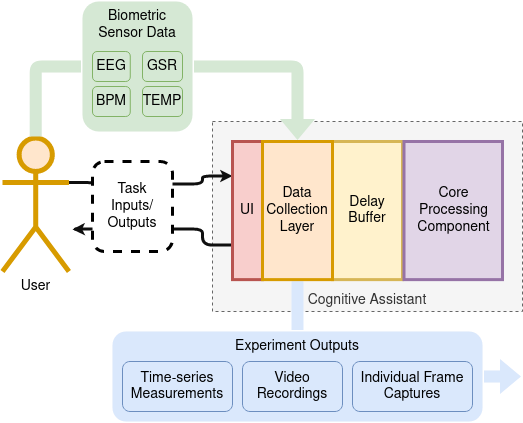}
  \caption{Experimental test-bed.
  Participants interact with the cognitive assistant through task-related inputs and outputs --- in practice, these correspond to the video feed captured by the assistant and the instructions provided by it.
  The assistant itself has been instrumented with a data collection layer, which collects and processes experiment-related data such as biometric signals from the participants (these are merely processed here and do not form part of the inputs to the cognitive assistant as such, however), and a delay buffer, which introduces controlled delays in the transit of information from the core processing component.}%
  \label{fig:experimentaltestbed}
\end{figure}

The core elements of our experiment are shown in \cref{fig:experimentaltestbed}:
\begin{itemize}
    \item Subjects interact with a WCA while wearing an array of biometric sensors.
    \item The \emph{responsiveness} of the application, i.e.\ the interval of time between an input being provided to the system and the associated output returned to the user, is manipulated in real time. 
    The effects of these manipulations on the subjects are recorded and subsequently analyzed.    
\end{itemize}

This study was conducted with the approval of the Carnegie Mellon University Institutional Review Board.
Subjects were recruited from a pool of undergraduate students fulfilling a course requirement at Carnegie Mellon University. 
No particular exclusion criteria were applied. 
In total, 40 participants were recruited, all of them of college student age (\numrange{18}{25} years old).

\subsection{The Cognitive Assistance Application}

We used a modified version of the LEGO Assistant application introduced by \textcite{Chen:EarlyImplementation}.
This application belongs to a category of \emph{WCA}s that are designed to guide users through the execution of a sequential task.
Such applications constitute ``\emph{conversational computing tasks}'' in the taxonomy proposed by \textcite{dabrowsky:2011:40years}.
A set of instructions are to be performed by the user in a semi-predetermined order.
The results of the user performing these instructions are provided as inputs to the system.
These inputs may either be correct, in which case the system proceeds to output the next instruction, or incorrect, in which case a procedurally generated corrective step that fixes the mistake is provided to the user.

\begin{figure*}[h]
  \centering
  \includegraphics[width=.8\textwidth]{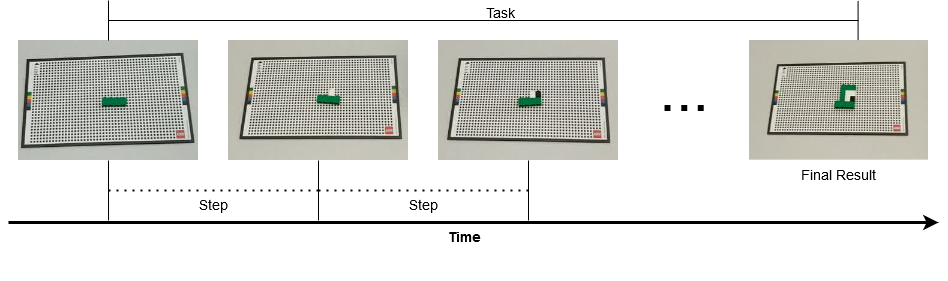}
  \caption{Example of a Cognitive Assistance task and its component subtasks and steps.}\label{fig:task:steps}
\end{figure*}

In more formal terms, we can provide definitions for \emph{task}, \emph{subtask}, and \emph{step} in such an application as:
\theoremstyle{definition}
\begin{definition}[Task and subtask]
  A \emph{task} will be understood as a finite sequence of instructions to be performed in order.
  \emph{Subtask} will refer to a specific action to be performed by the user, described by a single instruction.  
  See for instance \cref{fig:task:steps}, which pictures the task of assembling a simple LEGO model, with each subtask corresponding to the addition of a specific LEGO piece to the current model.  
\end{definition}

\begin{definition}[Step]
  Interval of time delimited by two consecutive instructions, see \cref{fig:task:steps,fig:cogassist:step}.
  Steps are characterized by the actions of the user and the assistant.
  At the beginning of a step, an instruction is given to the user.
  The user then proceeds to perform the subtask specified by the instruction, while the cognitive assistant is continuously sampling the subtask state at specified intervals. 
  While the subtask remains unfinished, the results of the processing of the sampled inputs are discarded.
  Once the user finishes the required action, the next sample taken will contain a valid input, and thus the cognitive assistant will provide a new instruction. 
  This finishes the current step and potentially begins a new one if there remain instructions to be performed in the task.
\end{definition}

In the base LEGO Assistant, the task consists of steps leading the user through the assembly of a LEGO model; each subtask requires the user to append a LEGO brick to the model at a specific location and orientation.
The system monitors progress through a video feed and provides timely feedback in the form of visual and textual instructions to guide the user towards the desired end result. 

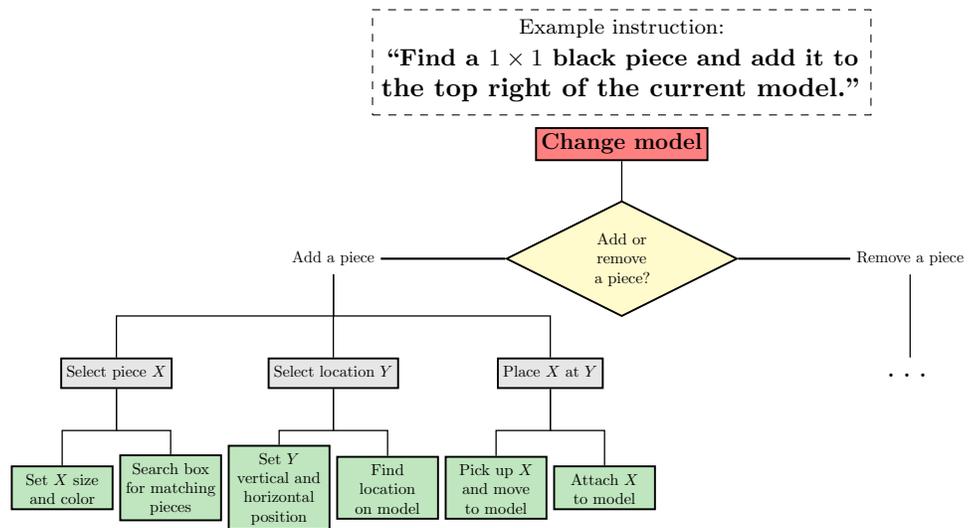
\begin{figure*}[h]
  \centering
  \begin{tikzpicture}[grow=down, scale=.6]
    \tikzset{edge from parent/.style= 
            {draw, thick, edge from parent fork down},
            every tree node/.style=
            {draw,very thick,align=center,scale=1,fill=black!10!white},
            every leaf node/.style=
            {fill=green!60!black!25!white,text width=2cm,minimum width=2cm},
            level distance=1in,sibling distance=.05in,
            execute at begin node=\strut
            }

    \node[draw,dashed,align=center](command){
        \footnotesize{Example instruction:}\\
        \small\textbf{``Find a $1 \times 1$ black piece and add it to}\\
        \textbf{the top right of the current model.''}
    };

    \begin{scope}[shift={($(command.south)+(0,-.25in)$)}]
        \Tree [.\node[fill=red!50!white]{\textbf{\Large{Change model}}};
                \node[diamond, aspect=2, fill=yellow!25!white](choice){Add or remove a piece?};
                ]
    \end{scope}

    \begin{scope}[shift={($(choice.west)-(1.5in,0)$)}]
        \Tree [.\node[draw=none, fill=none](add){Add a piece};
                [.{Select piece $X$} 
                    {Set $X$ size and color}
                    {Search box for matching pieces} ]
                [.{Select location $Y$} 
                    {Set $Y$ vertical and horizontal position}
                    {Find location on model} ]
                [.{Place $X$ at $Y$} 
                    {Pick up $X$ and move to model}
                    {Attach $X$ to model} ]
            ]
    \end{scope}

    \begin{scope}[shift={($(choice.east)+(1.5in,0)$)}]
        \Tree [.\node[draw=none, fill=none](remove){Remove a piece};
                \node[draw=none, fill=none]{\Huge{\ldots}}; 
            ]
    \end{scope}

    \draw [thick] (choice.west) -- (add.east);
    \draw [thick] (choice.east) -- (remove.west);

\end{tikzpicture}%
  \caption{Hierarchical cognitive structure of a step in the LEGO task.}
  \label{fig:lego:hierarchical}
\end{figure*}

The LEGO assistant has features that make it a good target for assessment of delay effects in a cognitive assistant:
\begin{enumerate}
    \item \ It is easily understood by users and requires essentially no training. 
    \item\ The step-by-step nature of these applications simplifies the isolation of the relevant experimental variables and the effects of the delay.
    \item The states of the display can be recognized by simple image processing algorithms and do not require extensive training as for machine-learning-based applications.
    \item Each step has an intrinsic hierarchical cognitive structure (see \cref{fig:lego:hierarchical}), affording multiple levels of cognitive control. 
\end{enumerate}

The original design of the LEGO assistance application was based on a client-server model communicating over a wireless network, with the client software running on a wearable device and the server software deployed on a cloudlet. 
For the purposes of this study, this design was altered to be executable on a single, non-networked computer, in order to eliminate the stochastic effects of jitter and latency on the network link.
By this means, we achieve much more fine-grained control over the latencies to which the system is subject.
Additionally, this greatly simplified the instrumentation of the application.
Instructions were output in image and text form to a computer display situated on a table directly in front of the participants.
Participants performed the instructions on the table, these actions being captured by a high-definition camera located on top of the display.

Finally, a data collection and experimental layer was implemented between the user interface and video capture and the core processing component of the LEGO Assistant.
This layer controlled the manipulated experimental variables and recorded the measured biometric and task-related effects.

\subsection{The Experimental LEGO Task}

\begin{figure}[hp]
  \centering
  \begin{subfigure}{\textwidth}
    \centering
    \includegraphics[width=.8\textwidth]{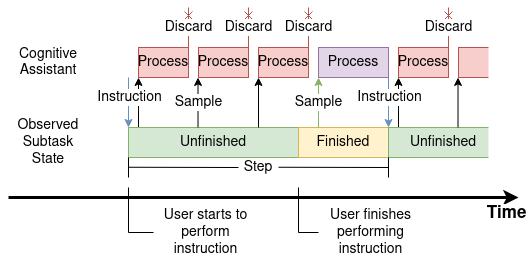}
    \caption{Structure of a step in a generic cognitive assistance application. 
    The assistant provides an instruction to the user and continuously samples the subtask state; inputs captured while the subtask is unfinished are silently ``discarded'' (i.e.\ they do not cause the generation of a new instruction) by the assistant, as they do not contain relevant information.
    However, once the user finishes performing the given instruction, the next input \emph{will} cause the generation of a new instruction, which will subsequently be provided to the user.}%
    \label{fig:cogassist:step}
  \end{subfigure}
  \medskip%
  \begin{subfigure}{\textwidth}
    \centering
    \includegraphics[width=.8\textwidth]{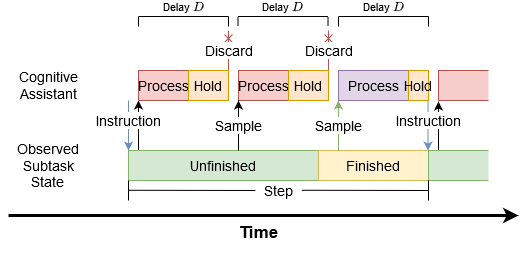}
    \caption{Modified structure of a step in the experimental task. In contrast to \cref{fig:cogassist:step}, an additional variable segment of time is introduced immediately following the processing of the input frame in the cognitive assistant, in order to extend the perceived processing time of the input to a specific target delay.}%
    \label{fig:cogassist:step:delay}
  \end{subfigure}
  \medskip%
  \begin{subfigure}{\textwidth}
    \centering
    \includegraphics[width=.8\textwidth]{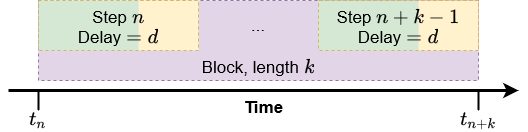}
    \caption{Structure of a block in the experimental task.}%
    \label{fig:cogassist:block}
  \end{subfigure}%
  \caption{Components of the cognitive assistance task.}
\end{figure}

For the experimental LEGO task, a key modification was made to the structure of the steps.
After each the processing of each input frame is completed, the result is withheld for a variable period of time until a specific target \emph{delay} is reached, as illustrated in figure \cref{fig:cogassist:step:delay}.
The length of this delay is one of two independent variables we manipulated for the experiment.
Seven levels of delay were used --- no added delay (which we will also refer to as \SI{0}{\second} delay), \SIlist{0.6;1.125;1.65;2.175;2.7;3.0}{\second} --- chosen based on the latency bounds found previously by \textcite{Chen:AnEmpiricalStudyOfLatency}.
A value of  \SI{600}{\milli\second} was identified as the bound where users start noticing delays in the assistant.
Conversely, \SI{2.7}{\second} was identified as the upper bound on delays after which the application is considered to be in such a degraded state it is basically ``unusable''.
Thus, our selection of delays is centered around the range of delays where latency is noticeable to users, but the application remains in a ``usable'' state, while including one delay value in the unnoticeable range and one completely in the ``unusable'' range.

Additionally, in order to study the effects of a delay applied across multiple steps, we implemented an experimental design component called a \emph{block}:

\begin{definition}[Block]
  A sequence of consecutive steps within a task subject to the same delay; see \cref{fig:cogassist:block}.
  The \emph{length} of a block corresponds to the number of steps it encompasses, and is the second of the two independent variables manipulated in our experiment.
  We used values of \numlist{4;8;12} steps for the lengths, values chosen as representative of the number of steps in tasks in the base LEGO Assistant application.
  Additionally, we define the \emph{duration} of a block, as the time elapsed between the start timestamp of the first step in the block and the end timestamp of the final step in the block; e.g.\ for \cref{fig:cogassist:block}, the duration of the pictured block would be \( t_{n+k} - t_{n} \).
\end{definition}

To assign tasks to the participants, a pseudo-random permutation of the combinations of block length and delay was first generated, and a unique sequence of steps assigned to each of these combinations to create 21 unique blocks.
A \emph{Latin square}-type design was then used to reorder this initial permutation in order to generate a task for each participant.
This ensures a counterbalance of the order of the blocks across participants and avoids systematic learning effects. 
The design rotates the block types (as defined by length and delay) across participants so that each type is tested in each ordinal position, but it only coarsely samples from the \( 21 \times 20 \) possible sequences from one block type to another.
Note that unlike the base LEGO assistant task, in which instructions led participants through the assembly of a specific model, the experimental LEGO task consisted of a sequence of instructions with no evident goal.
Users were directed to either add a piece to the ongoing model or to remove a piece, and blocks were designed in such as way so as to ensure that the transitions between them were invisible to the user. 

In this paper we will thus consider blocks to be our basic element of study, and most aggregations will be done at this level (with a few exceptions).
For this, we will need additional definitions:

\begin{definition}[Delay associated with a block]
  We will refer to the delay of a block as the delay applied to every frame of every step in that block.
\end{definition}

\begin{definition}[Execution time]

  Given the variability in the system latency to detect step completion, correcting step-completion time for system response time by a fixed amount would not enable sufficient precision.
  We therefore estimated the \emph{execution time} empirically for an individual step as the total time between the user receiving the instruction for a subtask and their presenting the completed subtask to the system.


  Formally, for an arbitrary step, we define the sequence \( S = \{ t_{0}, t_{1}, \ldots, t_{n} \} \) as the sequence of timestamps corresponding to the sampling instants during the step.
  \( t_{0} \) corresponds then to the instant where the instruction for the step is given to the user (and the first sample is taken), and \( t_{n} \) to the timestamp of the last sample before a new instruction is given (or conversely, the sample which captured the finished subtask as presented by the user).
  If we define \( t_{c} \) as the instant marking when the user finished and presented the task to the system, we infer the following:

  \begin{itemize}
    \item \( t_{c} < t_{n} \), since by definition the user must have finished the task before the system took the sample that triggered a new instruction.
    \item \( t_{c} > t_{n - 1} \), since otherwise the system would have triggered a new instruction after some \( t_{k}, k \in [0, n-1] \) instead of after \( t_{n} \).
  \end{itemize}

  Therefore, \( t_{n-1} < t_{c} < t_{n} \) (see also \cref{fig:exectime:diagram}).
  Due to the discrete sampling of the task state, there remains some imprecision in the estimate of execution time relative to \( t_{c} \).
  However, this introduces no bias in the results, as we can infer that \( t_{c} \) is uniformly distributed in the range \( (t_{n-1}, t_{n}) \).
  We therefore calculate execution time for each step as \( t_{c} - t_{0},\; t_{c} \in U(t_{n-1}, t_{n}) \), which on average works out to an adjustment of the observed time by 1.5 times the mean sampling rate of the step.
  This is also aggregated into an average at the block level.

  \begin{figure}
    \centering
    \includegraphics[width=.85\textwidth]{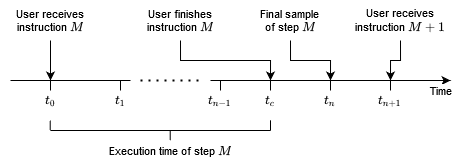}
    \caption{Visualization of the execution time of a step.}%
    \label{fig:exectime:diagram}
  \end{figure}

\end{definition}

\subsection{Collected Data}

The collected data from the experiments fall into four categories: behavioral and personality indicators, \emph{frame-to-frame} metrics, biometric data, and video recordings.

\subsubsection{Behavioral and Personality Indicators}

Before beginning the experimental procedure, participants were asked to fill out two questionnaires.

The first of these, the Big Five Inventory of Personality (\emph{BFI}~\autocite{john1999:bfi}), consists of 44 questions to be answered on a 5-point \emph{Likert}-type scale, assessing the traits of \emph{agreeableness: detached to compassionate} (8 questions), \emph{conscientiousness: careless to organized} (9 questions), \emph{extroversion: reserved to outgoing} (8 questions), \emph{neuroticism: secure to sensitive} (8 questions), and \emph{openness: cautious to curious} (9 questions).
Of these, extroversion and neuroticism have been related to tolerance for delay~\autocite{hirsh2008delay}. 

The second survey, the Immersive Tendencies Questionnaire (\emph{ITQ}~\autocite{witmer1998:itq}), comprises 29 questions, 28 of which were used for the study (one categorical question was disregarded), assessing the sub-scales of \emph{involvement}, the tendency to become involved in activities; \emph{focus}, the tendency to maintain focus on current activities, and \emph{games}, the tendency to play games.
These questions use a 7-point horizontal scale  with opposing descriptors anchoring  the ends.
Participants were asked to mark the appropriate point in the scale, and these responses were converted to a numerical value between 1 and 7 for processing. 

In post-processing, the obtained scores for both questionnaires were normalized to fall in the \( [0, 1] \) range for ease of interpretation. 
See \cref{tab:qscores} for their means and standard deviations.

\begin{table}[h]
  \centering
  \caption{Means and standard deviations of normalized questionnaire scores.}\label{tab:qscores}
  \setlength{\tabcolsep}{0pt} 
  \begin{tabular*}{\columnwidth}{@{\extracolsep{\fill}\quad}clrr@{}}
    \toprule
    \textbf{Questionnaire} & \textbf{Metric} & \textbf{Mean} & \textbf{SD}\\
    \midrule
    \multirow{5}{*}{BFI} 
        & Agreeableness &               0.705 &                  0.136 \\
        & Conscientiousness &               0.562 &                  0.149 \\
        & Extroversion &               0.491 &                  0.180 \\
        & Neuroticism &               0.524 &                  0.230 \\
        & Openness &               0.677 &                  0.152 \\
    \cline{1-4}
    \multirow{4}{*}{ITQ} 
        & Focus &               0.626 &                  0.126 \\
        & Games &               0.463 &                  0.261 \\
        & Involvement &               0.568 &                  0.175 \\
        & Total &               0.540 &                  0.092 \\
    \bottomrule
  \end{tabular*}
\end{table}

\subsubsection{Frame-to-frame metrics}
During the execution of the task, we logged every event occurring in the application pipeline.
Each incoming frame (including frames discarded by the assistant), as well as its associated outputs, was logged at multiple points in the process along with associated metadata such as currently implemented delay as specified by the experimental design.
This allowed us to extract metrics relating to the performance of the task, such as the time spent by participants on particular steps, any mistakes made, etc.
In particular, it made possible the easy segmentation of the other time-series data we collected into our main unit of analysis, the aforementioned \emph{block}.

\subsubsection{Biometric Data}
The participant wore devices to acquire four physiological measures:

\begin{itemize}
  \item galvanic skin response (\emph{GSR});
  \item accelerometer data from the dominant wrist;
  \item brain activity in the form of electroencephalography (\emph{EEG});
  \item and heart rate.
\end{itemize}

These metrics have been used as indicators of stress and cognitive load by an ample body of previous research~\autocite{khawadi2015:usinggsrtrust,kuikkaniemi2010:biofeedback,solovey2014:classifyingdriverworkload}.
More specifically, galvanic skin response (\emph{GSR}, also known as electrodermal activity) is the measure of the variation of the conductive properties of human skin due to changes in the state of the sweat glands.
It is interpreted as an indicator of physiological arousal and has long been a widely used metric in studies seeking to characterize mental workload~\autocite{peterson1907psycho,Healey2005,Son2010,khawadi2015:usinggsrtrust,kuikkaniemi2010:biofeedback,solovey2014:classifyingdriverworkload,}.
Electroencephalography (\emph{EEG}) refers to the monitoring of brain activity through the measurement of the fluctuations of the electric field surrounding the brain.
\emph{EEG} measures the voltage fluctuations due to electrical activity within neurons in the brain, which result in distinct waves of specific frequencies associated with different contexts, emotions and actions.
\emph{EEG} has  been used to measure cognitive load in the context of human-computer interactions~\autocite{Antonenko2010,Grimes2008,kumar2016measurement}.

Wrist acceleration, GSR and heart rate data were obtained using the \textcite{empatica:e4} biosensing wristband.
Accelerometer data was sampled at \SI{32}{\hertz},
GSR was sampled at \SI{4}{\hertz}, and instantaneous heart rate was calculated from a \emph{blood volume pulse} (BVP) signal sampled at \SI{64}{\hertz}.
Participants were asked to wear the device for approximately 10--15 minutes before starting the experiment, in order to allow the sensors to reach an stable equilibrium and establish a baseline for the signals.

The E4 wristband was chosen due to its small, non-invasive and wireless form factor (samples were streamed to the system over Bluetooth LE) and for the fact that its use in research has been experimentally validated in previous research~\autocite{ragot2017emotion, mccarthy2016validation}.
The E4 also includes a skin temperature thermometer; however, due to failure to reach equilibrium, the measure was not used for the present study.

For the EEG data we employed the \textcite{openbci:headbandkit}, a kit consisting of a number of dry electrodes fastened to a Velcro headband.
It thus provides a quick, easy and non-invasive way of obtaining EEG signals from participants.
Electrodes were placed according to the \emph{10--20 Electrode System}~\autocite{eeg1020system:1961} on the \emph{Fp1} and \emph{Fp2} points, in order to capture brain activity in the frontal lobe.
Ground and reference electrodes were positioned on the right and left earlobes respectively.

This kit was paired with the \textcite{openbci:ganglion} for the actual acquisition of the signals, which were sampled at \SI{200}{\hertz} and streamed over Bluetooth to the experiment computer.

Following capture, the  EEG signal was post-processed in the following fashion:
\begin{enumerate}
  \item Since our main interest was in the \( \alpha \) (\SIrange{8}{12}{\hertz}) and \( \beta \) (\SIrange{12}{30}{\hertz}) bands, a low-pass Butterworth filter of order 8 and cutoff frequency \SI{40}{\hertz} was applied to filter out noise in the higher end of the spectrum (in particular, noise from the board power supply at \SI{60}{\hertz}).
  \item A high-pass Butterworth filter of order 8 and cutoff frequency \SI{0.1}{\hertz} was then applied to filter out noise at the low end of the spectrum.
  \item Subsequently, a pair of cascaded Savitsky-Golay filters~\autocite{savitzky1964smoothing}, both of order 8 and window size 21, were applied to the signal in order to smooth out noise, as proposed by \textcite{agarwal2017eeg}.
  \item A spectrogram was then calculated for the signal.
  \item Finally, the power over time for each EEG frequency band is obtained by integrating over the relevant frequencies for each time-step in the spectrogram.
\end{enumerate}

\begin{figure}[h]
  \centering
  \includegraphics[width=.8\textwidth]{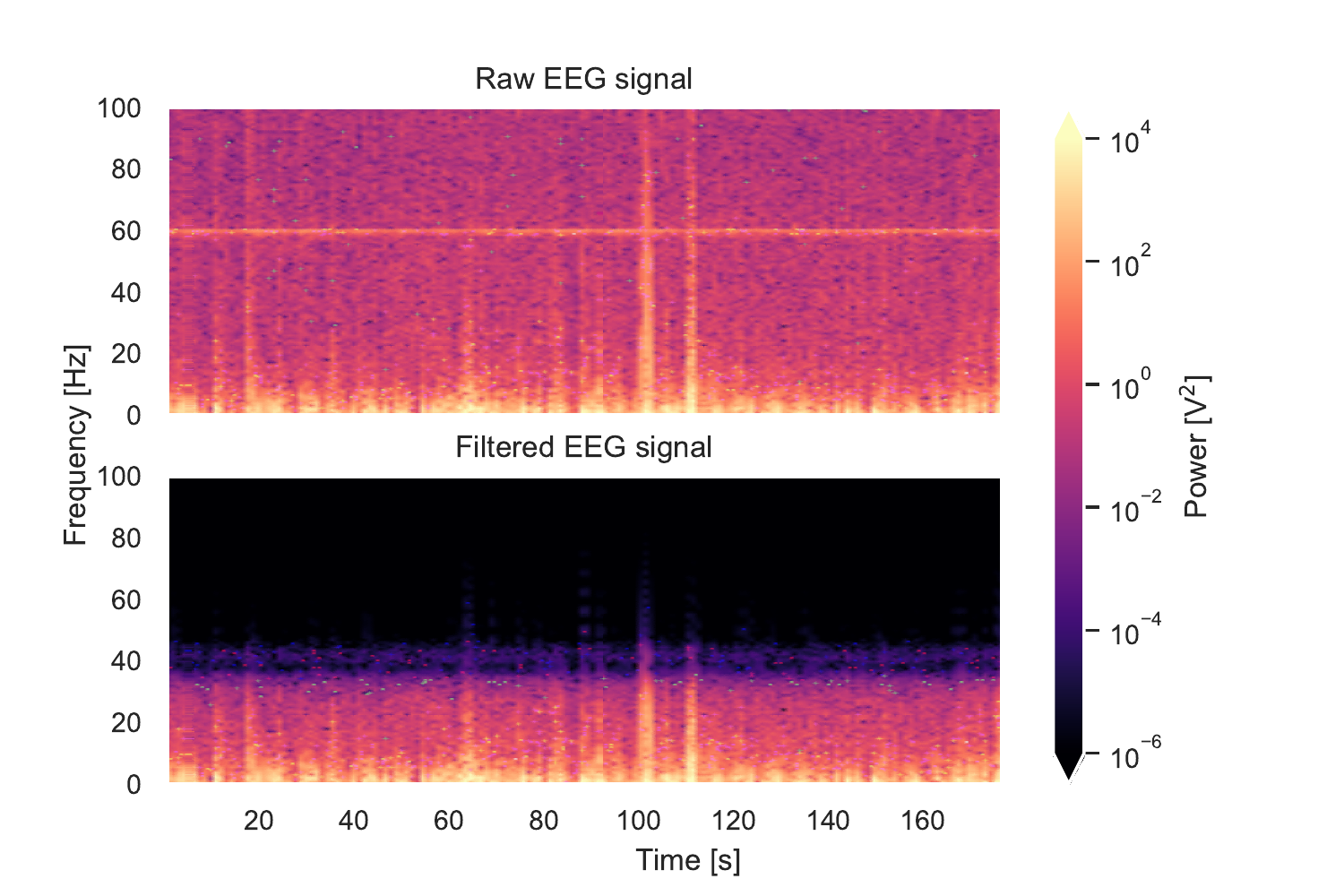}
  \caption{Comparison of an EEG signal before and after filtering.
  Note in particular the artifact from the DC power supply of the capture board around \SI{60}{\hertz} in the raw signal, and how it essentially disappears after filtering.}%
  \label{fig:eeg:raw_vs_filt}
\end{figure}

An example of the effects of the filtering on the raw EEG signal can be observed in \cref{fig:eeg:raw_vs_filt}.

\subsubsection{Video recordings}

During the task, participants were recorded by two separate cameras.
One camera was angled downwards, towards the table, the LEGO board, and the participant's hands.
This camera was used to capture the necessary inputs for the LEGO assembly task as well as to record the actions performed by the user.
The second camera was angled horizontally, towards the participant, in order to record their facial expressions during the execution of the task.

\begin{figure}[h]
  \centering
  \begin{subfigure}[t]{.35\textwidth}
    \centering
    \includegraphics[height=140pt]{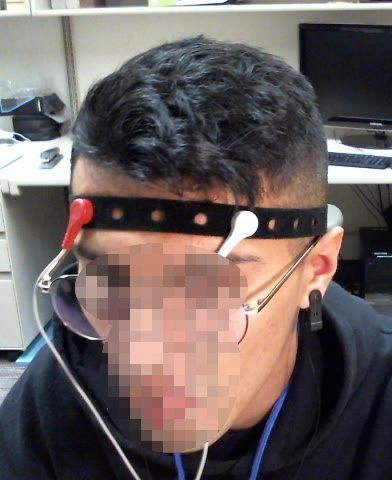}
    \caption{Frame from the face recording of a random participant, clearly showing the locations of the EEG electrodes.}
  \end{subfigure}%
  \hfill%
  \begin{subfigure}[t]{.60\textwidth}
    \centering
    \includegraphics[height=140pt]{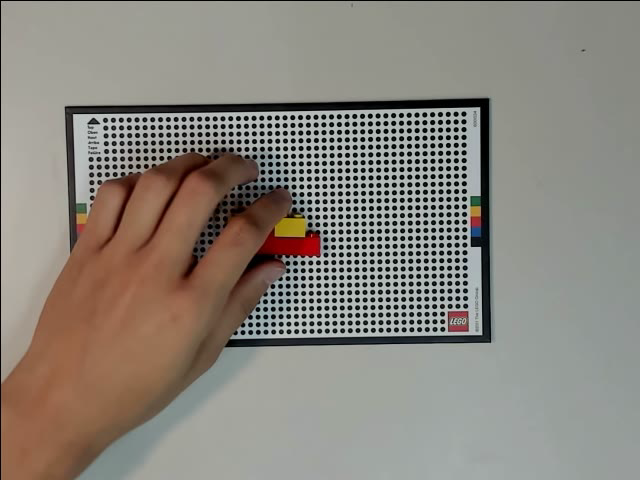}
    \caption{Frame from the board recording of the same participant.}
  \end{subfigure}
  \caption{Sample frames captured during the experiment.}\label{fig:sampleframes}
\end{figure}

Both video feeds were captured at a rate of \num{24}~FPS (i.e.\ with a sampling interval of \( \SI[parse-numbers=false,per-mode=symbol]{41.\overline{6}}{\milli\second} \)) in parallel processes to ensure a constant rate of capture.
Examples of the captured frames can be seen in \cref{fig:sampleframes}.
The video feeds were not used for the present study, but may be utilized in future analysis.

\section{Results}\label{sec:results}

The results consist of execution time per step. and the outcome of physiological variables: heart rate, GSR, and EEG.\@
Each will be discussed in turn.

\subsection{Execution Time}


Before describing the analysis and results relating to execution time, it should be noted that participants' performance during the execution of the task was error free; all steps were completed as instructed. 

\begin{figure}[h]
    \centering
    \includegraphics[width=.8\textwidth]{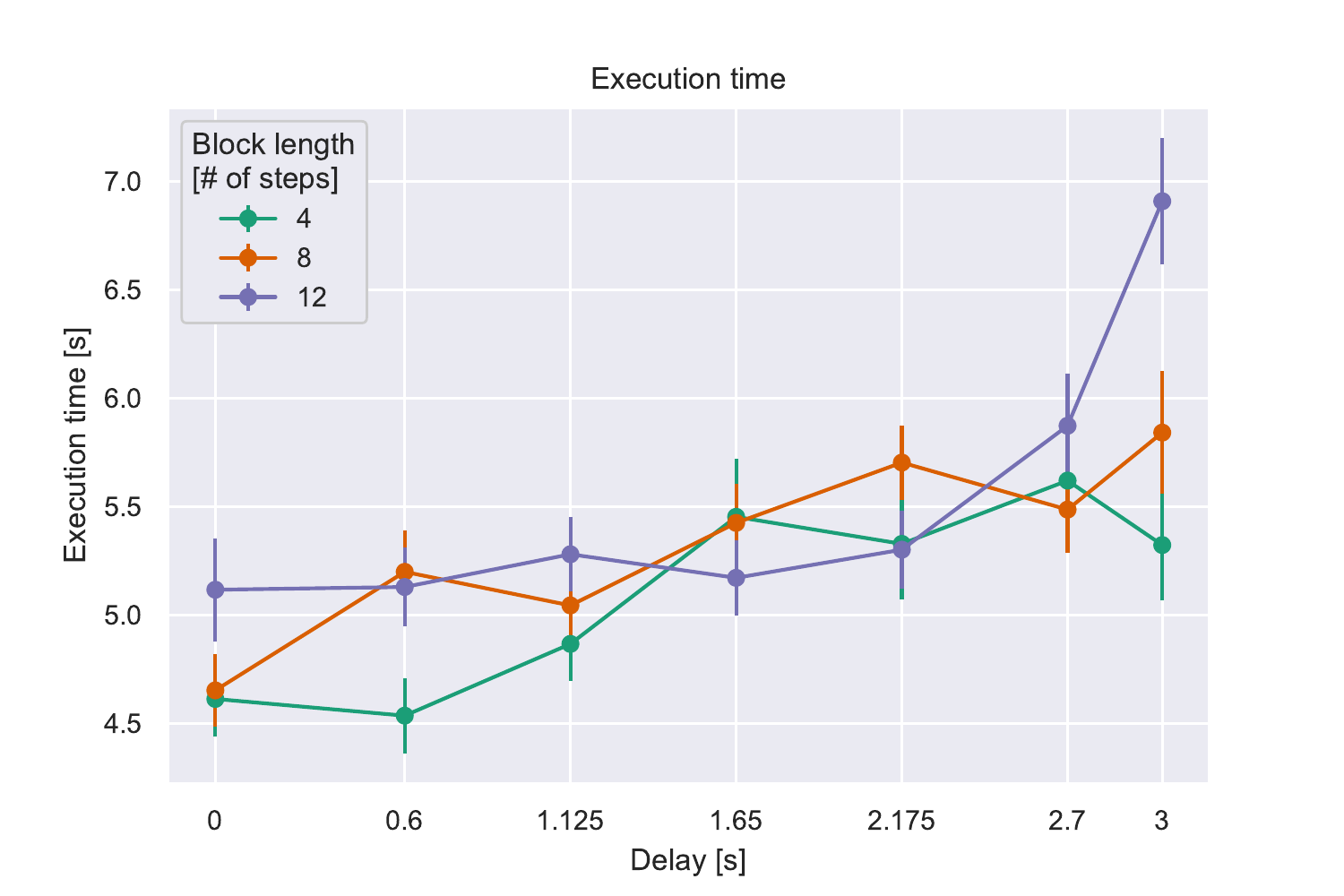}
    \caption{Per-step execution time by block length vs.\ delay. Error bars indicate the Standard Error of the Mean (S.E.M.)}\label{fig:exectime}%
\end{figure}

\cref{fig:exectime} shows the mean per-step execution time per block, averaged over block length (number of steps) and artificial delay.
We can clearly see a trend for the execution time to increase with the delay, increasingly so for longer blocks. 
Since the per-step execution time  compensates for the added delay in the measure \emph{per se}, this trend must result from the participants' behavioral adjustment to the delay.
This leads to one of the key outcomes from this study: 
\emph{participants tend to act more slowly on steps affected by longer delays} --- i.e.\ there is evidence of a pacing effect in users' behavior with respect to the responsiveness of the system.

We confirmed this effect through an \emph{analysis of variance} (ANOVA) with factors of block length and delay.
An ANOVA test uses the \emph{F-test} statistic, which is a ratio of the variability in the data introduced by experimental manipulations (and their interactions) to the variability in the data attributable to randomness.
The degrees of freedom represent the number of observations going into each of these variability measures.
As we are using within-subject designs, the variability in the data due to randomness is estimated by the differences among subjects with respect to the effects of the experimental variables.
The partial-\( \eta^2 \) (\(\eta_{p}^{2}\)) statistic is a measure of effect size, corresponding to the proportion of variance explained by the effect after excluding variance from others, and \( p \)-value, a measure of the probability of the effect under the null hypothesis.
Conventionally \( p < .05 \) is the criterion for significance.
The current length X delay ANOVA found significant main effects of both factors and the interaction, shown in \cref{tab:anova:exectime}.

\begin{table}[h]
  \centering
  \caption{Significant effects on per-step execution time from ANOVA on factors delay and block length.}\label{tab:anova:exectime}
  \setlength{\tabcolsep}{0pt}%
  \begin{tabular*}{\columnwidth}{@{\extracolsep{\fill}\quad}lrrr@{}}
    \toprule
    \textbf{Factor} & \textbf{F-test} & \( p \) & \( \eta^{2}_{p} \) \\
    \midrule
    length         &    \(F(2, 78) = 9.59\) &   \(< 0.001\) &           \(0.20\) \\
    delay          &  \(F(6, 234) = 15.52\) &  \(< 0.0001\) &           \(0.28\) \\
    length \(\times\) delay &  \(F(12, 468) = 3.84\) &  \(< 0.0001\) &           \(0.09\) \\
    \bottomrule
  \end{tabular*}
\end{table}

\begin{figure}[h]
    \centering
    \includegraphics[width=.8\textwidth]{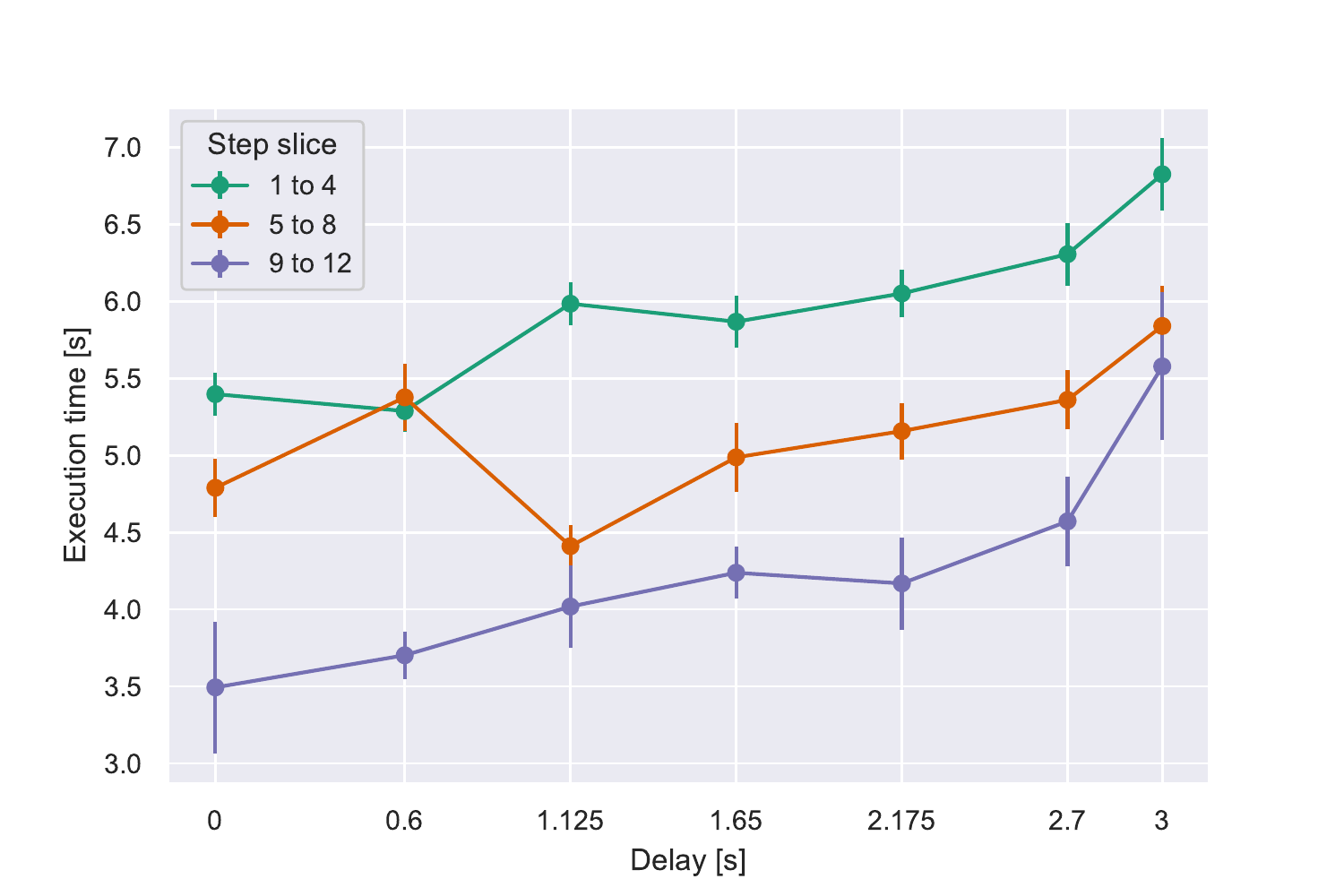}
    \caption{Mean per-step execution time vs.\ delay, by step slice.
    Error bars indicate S.E.M.}
    \label{fig:exectime:delay:slice}
\end{figure}

Further analysis focused on the progressive effect on per-step execution time as additional steps occurred at a constant delay.
For this purpose, the steps within a block were aggregated over sequences of 4, constituting a \emph{slice}.
Note that the first slice within a block is procedurally identical for all block lengths, in the sense that a participant currently performing a step in the first slice of a block has no way of predicting if the block ends after step 4 or not.
The same logic can be applied to slice 2 for blocks of length 8 and 12.
Accordingly for each participant, slice 1 data (steps 1--4) were pooled over all 3 block lengths, slice 2 (steps 5--8) were combined for block lengths 8 and 12, and slice 3 comprised the last four steps in blocks of length 12.
An ANOVA on slice number (1--3) and delay (7 values) yielded effects of slice number, delay, and the interaction, detailed in \cref{tab:anova:exectime:slice}.
As shown in \cref{fig:exectime:delay:slice}, blocks with shorter delays showed a trend for the execution time to progressively decline over the course of the block (i.e., by slice number), indicating that the participant accommodated to the feedback pace with more efficiently timed responses.
With the longest delays, where the execution time per step was longest, the slow-down persisted; that is, the system response time hindered the participant's execution throughout the course of the block.

\begin{table}[h]
  \centering
  \caption{Significant effects on per-step execution time from ANOVA on factors block slice and delay.}\label{tab:anova:exectime:slice}
  \setlength{\tabcolsep}{0pt} 
  \begin{tabular*}{\columnwidth}{@{\extracolsep{\fill}\quad}lrrr@{}}
    \toprule
    \textbf{Factor} & \textbf{F-test} & \( p \) & \( \eta^{2}_{p} \) \\
    \midrule
    slice         &   \( F(2, 78) = 88.79 \) &  \( < 0.0001 \) &         \( 0.69 \) \\
    delay         &  \( F(6, 234) = 14.13 \) &  \( < 0.0001 \) &         \( 0.27 \) \\
    slice \( \times \) delay &  \( F(12, 468) = 2.49 \) &    \( < 0.01 \) &         \( 0.06 \) \\
    \bottomrule
  \end{tabular*}%
\end{table}

\begin{figure}[h]
  \centering
  \includegraphics[width=.8\textwidth]{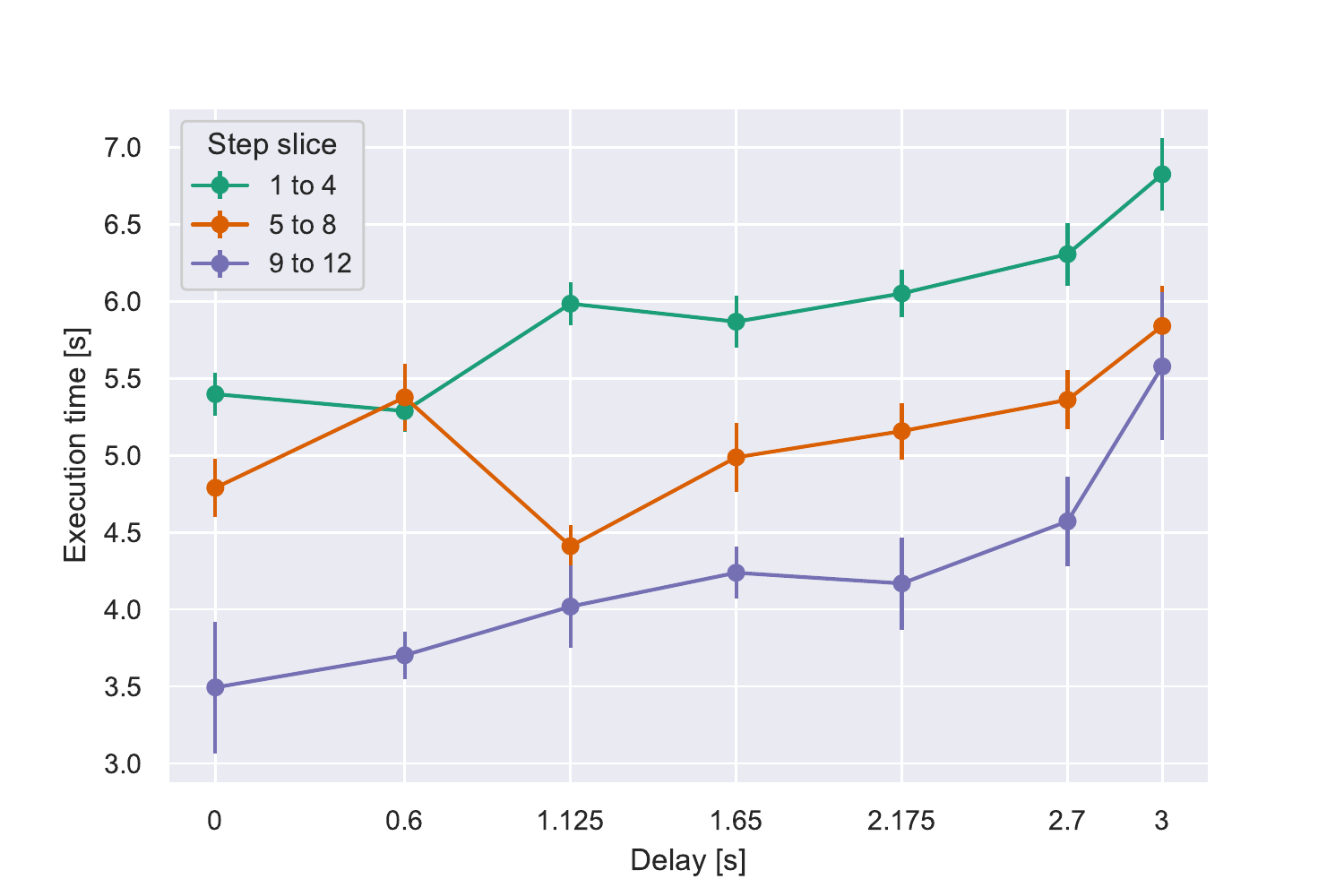}
  \caption{Per-step execution time across the first four steps after a block transition from block \( B_{k-1} \) to \( B_k \). Error bars indicate S.E.M.}\label{fig:exectime:transition}%
\end{figure}

The data also allowed us to perform sub-analyses to assess the effects of carryover from one delay to another.
Specifically, we measured the per-step execution time for the first four steps of a block when participants transferred from a relatively long delay (\SIrange{2.175}{3.0}{\second}) versus a short delay (\SIrange{0}{1.65}{\second}).
Note that use of the first four steps controls for block length.
We performed analyses where participants transitioned from either a short- or long-delay block into a
\begin{enumerate*}[label=(\roman*), before=\unskip{: }, itemjoin={{; }}, itemjoin*={{; and }}]
  \item no delay block (36 subjects)
  \item \SI{1.65}{\second} delay block (40 subjects)
  \item \SI{2.7}{\second} delay block (40 subjects)
  \item \SI{3.0}{\second} delay block (37 subjects).
\end{enumerate*}
The destination delays were chosen so as to maximize the number of participants which contributed samples to the analyses, the results of which are pictured in \cref{fig:exectime:transition}.
We found that transitions from long-delay blocks carried over to significantly increase the per-step execution time of initial steps in the destination block for three of the four transitioned-into delays that were assessed (\SIlist{0;2.7;3.0}{\second}, all \( p < .025\)).

In summary, we observe two direct effects on execution time due to added delays.
Firstly, we see a clear hampering of the improvement of execution time across steps that is otherwise evident across blocks.
Secondly, we notice that this effect lingers on even after the delay is removed, affecting subsequent blocks in the task.

\subsection{Acceleration}

\begin{figure}[h]
    \centering
    \includegraphics[width=.8\textwidth]{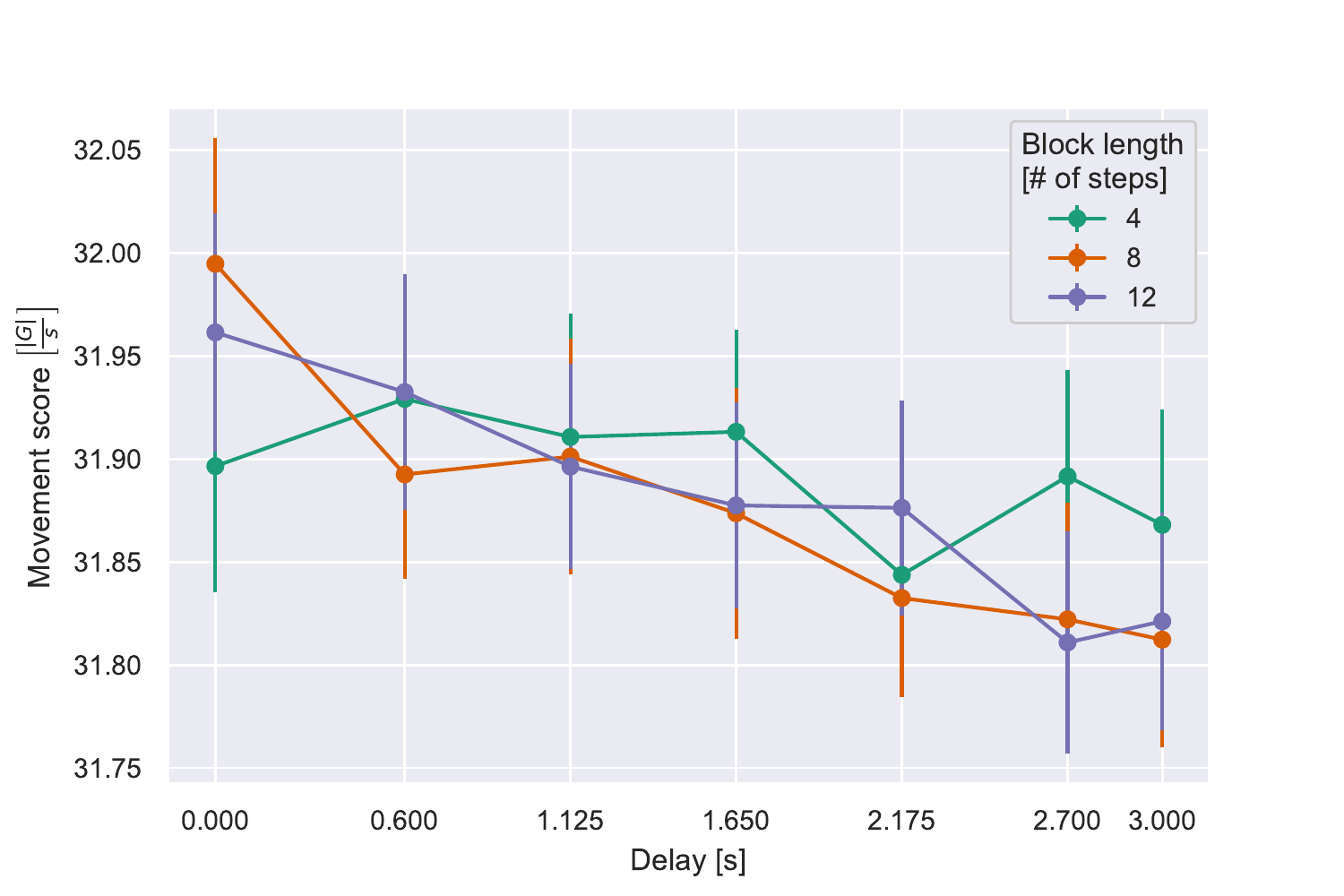}
    \caption{Movement score vs.\ delay, per block length. Error bars indicate S.E.M.}\label{fig:acc:delaylength}
\end{figure}

Acceleration data from the E4 wristband were taken in 3 axes defined relative to the device.
As our interest was in the amount of movement rather than the direction in space, we calculated a ``movement score'' for each block which we defined as:
\begin{equation}
    M_B = \frac{ \sum_{A_B} |\: \overrightarrow{\alpha_j} \:| }{\Delta T_{B}}
\end{equation}

where \( B \) is an arbitrary block, and \(A_{B} = \{ \overrightarrow{\alpha_0}, \overrightarrow{\alpha_1}, \ldots, \overrightarrow{\alpha_k} \} \) represents the set of acceleration vector samples collected for said block.
This score would include the time imposed by the delay and the time to decode the instructions and respond by moving the pieces so that the next state was recognized.
It is only during the last part of the step that explicit movement is required, so any additional accelerations would derive from arbitrary movements while processing and waiting.
We normalized the sum of the accelerations by the duration of the step to correct for differences in delay and the execution time, which tends to increase with delay as described above.
An ANOVA on block length and delay showed a significant effect such that movement score decreased with delay, as shown in \cref{tab:anova:acc} and \cref{fig:acc:delaylength}.
There was also a significant delay by length interaction, reflecting that delay effects occurred particularly for the longer lengths.

\begin{table}[h]
  \centering
  \caption{Significant effects on accelerometer data from ANOVA on factors delay and block length.}\label{tab:anova:acc}
  \setlength{\tabcolsep}{0pt} 
  \begin{tabular*}{\columnwidth}{@{\extracolsep{\fill}\quad}lrrr@{}}
    \toprule
    \textbf{Factor} & \textbf{F-test} & \textbf{\(p\)} & \textbf{\(\eta^{2}_{p}\)} \\ 
    \midrule
    delay & \(F(6, 234) = 18.56\) & \(<0.001\) & \(0.32\) \\
    length \( \times \) delay & \(F(12, 468) = 3.04\) & \(<0.001\) & \(0.07\) \\ 
    \bottomrule
  \end{tabular*}%
\end{table}

\begin{figure}[h]
  \centering
  \includegraphics[width=.8\textwidth]{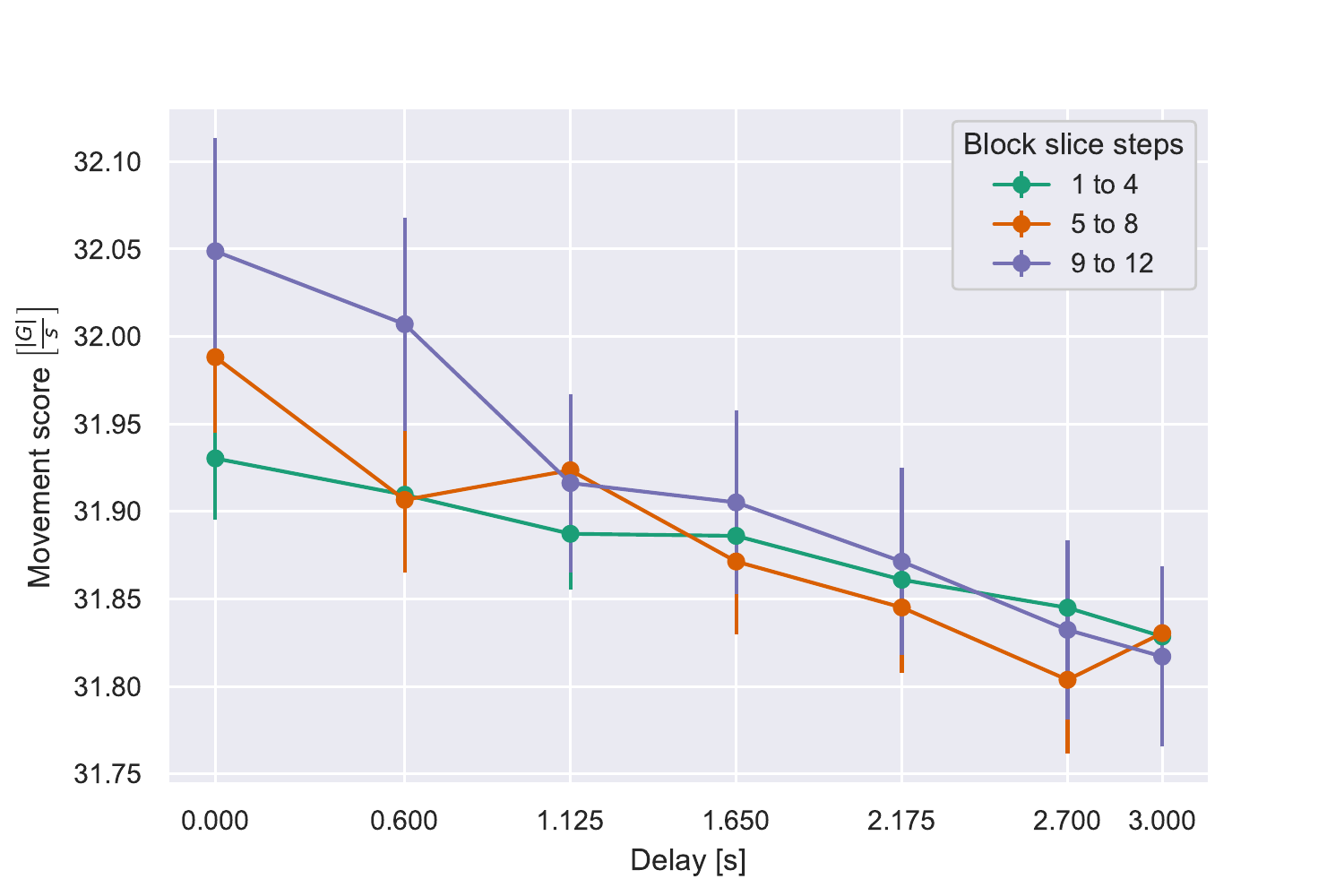}
  \caption{Movement score vs.\ delay, per block slice. Error bars indicate S.E.M.}\label{fig:acc:delayslice}
\end{figure}

We next conducted the same analysis as for execution time, dividing blocks into three ``slices'' of four steps each, such that slice 1 comprised the first four steps of all block lengths, slice 2 the second four steps of blocks of length 8 and 12, and slice 3 the last four steps of blocks 12 steps long.
As shown in \cref{fig:acc:delayslice} the effect of delay was attributable only to the later slices, yielding main effect of slice number and delay and an interaction (\cref{tab:anova:acc:slice}). 

\begin{table}[h]
  \centering
  \caption{Significant effects on accelerometer data from ANOVA on factors delay and slice number.}\label{tab:anova:acc:slice}
  \setlength{\tabcolsep}{0pt} 
  \begin{tabular*}{\columnwidth}{@{\extracolsep{\fill}\quad}lrrr@{}}
    \toprule
    \textbf{Factor} & \textbf{F-test} & \textbf{\(p\)} & \textbf{\(\eta^{2}_{p}\)} \\
    \midrule 
    slice & \(F(2, 78) = 8.39\) & \(<0.01\) & \(0.18\) \\
    delay & \(F(6, 234) = 26.92\) & \(<0.001\) & \(0.41\) \\
    delay \(\times\) slice & \(F(12, 468) = 4.08\) & \(<0.001\) & \(0.1\) \\
    \bottomrule
  \end{tabular*}%
\end{table}

These results present further evidence of the aforementioned pacing effect.
As a sequence of steps with a long delay unfolds, more of the step duration is spent without moving.
This can be interpreted in the context of the increased execution time at long delays demonstrated in the previous analysis.
Assuming that adding or deleting a LEGO block takes essentially the same amount of time and accelerates the wrist similarly at any one step, it appears that the participant simply remains stationary during the extra time that is induced by a series of long delays.
Accordingly, the acceleration per unit time, our movement score, is reduced.

\subsection{EEG}
Analyses were conducted on the log EEG power in the alpha band, beta band, and total of all bands measured.
Readings from the two frontally placed poles were highly correlated and were pooled into an average.
Logs were taken for analysis because the EEG distribution tended to have a rightward tail.
Twelve participants were excluded from the analysis, 9 due to device failure and 3 because of extreme values (i.e., the participant mean of log total power was greater than 3 s.d.s.\ from the mean of all participants).

The analyses then comprised 28 participants.\@
Omnibus ANOVAs were conducted with delay and block length (number of steps) as factors, on the EEG data from the alpha and beta band.
The analysis of alpha EEG found no significant effects.
Beta EEG showed only an effect of block length, \( F(2,54) = 3.56 \), \( p = .035 \), \( \eta^{2}_{p} = .12 \), reflecting a tendency for the 4-step length to produce lower log power (mean 3.9 vs. 4.1 for lengths of 8 and 12).
However, this effect was small and not consistent across delays.  

\begin{figure}[h]
    \centering
    \includegraphics[width=.8\textwidth]{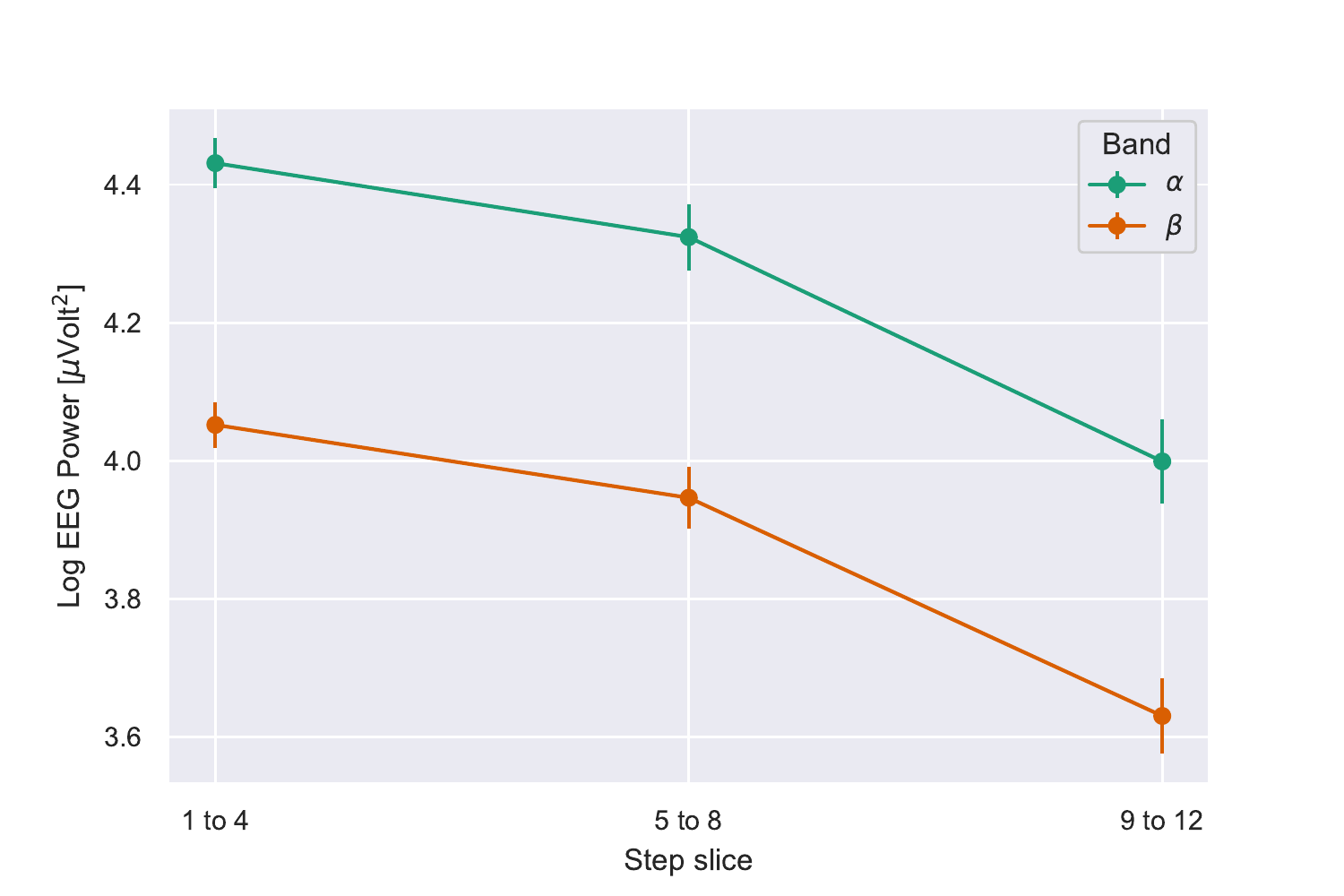}
    \caption{Mean Log EEG Power for alpha and beta bands per step slice. Error bars indicate S.E.M.}\label{fig:logeegpow:slices}
\end{figure}

Again the analysis dividing blocks into 4-step slices was conducted, with delay and slice number as factors.  
For both the alpha and beta bands, ANOVAs  yielded effects only of slice number --- see \cref{tab:anova:egg:alphabeta}.
Both bands showed the same tendency:  EEG declined as a sequence of steps with the same delay progressed.
These effects are shown in \cref{fig:logeegpow:slices}.
Thus, EEG taken from frontal locations mimics the execution time data in showing a decline over the course of a block, but unlike the execution time, there was no tendency for the decline in EEG to be reduced at longer delays.

\begin{table}[h]
  \centering
  \caption{Significant effects on log EEG power from ANOVA on factors delay and block slice.}\label{tab:anova:egg:alphabeta}
  \setlength{\tabcolsep}{0pt} 
  \begin{tabular*}{\columnwidth}{@{\extracolsep{\fill}\quad}llrrr}
    \toprule
    \textbf{Band} & \textbf{Factor} & \textbf{F-test} & \textbf{\(p\)} & \textbf{\( \eta^{2}_{p} \)} \\ \midrule
    alpha   & slice & \( F(2, 78) = 28.26 \) & \( <0.001 \) & \( 0.51 \) \\ 
    beta    & slice & \( F(2, 78) = 29.18 \) & \( <0.001 \) & \( 0.52 \) \\ 
    \bottomrule
  \end{tabular*}
\end{table}

\subsection{Galvanic Skin Response (GSR) and Heart Rate}
The measure of GSR was the log of total amplitude in the signal.
To control for effects of block length (steps~\( \times \)~delay) and the slow-down in execution time at longer delays, GSR data were normalized by the temporal duration of the block.
Due to sensor failure and the elimination of one subject with extreme values of GSR amplitude using the same rule as for EEG, the analysis comprised 34 participants.
No systematic trend due to delay or block length was observed.
In addition, the same analysis of slice position within block length as was performed for the EEG revealed no effects, indicating that GSR was stable across step positions within a block. 

Heart rate, measured in beats per minute, averaged \num{90.2}~BPM and showed no trend related to the experimental variables of delay and block length.
The same was true of the variability in beats per minute (average standard deviation across conditions was \num{20.3}).

The absence of systematic trends in both these results is interesting in the context of our initial suggestions of potential mechanisms relating delay in the application to human behavior.
In section \cref{ssec:potentialmechs} we proposed that emotional arousal during the execution of the task was a potential explanation for the effects observed in the human.
However, these results seem to refute this hypothesis, and will be further discussed in \cref{sec:discussion}.

\subsection{Individual Difference Analysis}

The Big 5 personality inventory and Immersive Tendencies Questionnaires (ITQ) were combined with outcome variables in an analysis of individual differences.
Given the previous results, we initially considered the following outcome variables:
\begin{itemize}
    \item execution time in the most demanding block (length 12, delay \SI{3.0}{\second}); 
    \item total of alpha and beta bands EEG in the first four steps at all lengths (``Slice 1'');
    \item average heart rate;
    \item and average log GSR.\@
\end{itemize}

Although the heart rate and GSR data had produced no significant effects in the analysis of experimental variables, they could in principle correlate with experimental outcomes across individuals. 
A principal-components analysis (\emph{PCA}) on a subset of these variables, shown in \cref{tab:pca}, was ultimately conducted on the 28 participants for which there was EEG data. 
Three of the personality measures were excluded after initial analyses indicated significant correlations among neuroticism, openness, agreeableness, and extroversion.
Neuroticism (with poles of sensitivity and security) was selected as most relevant to the issue of response to system delay and was included along with conscientiousness.
GSR was used in a subsidiary analysis, because only 25 participants had both EEG and GSR measures that were reliable.

\begin{table}[h]
  \centering
  \caption{Principal Component Analysis}\label{tab:pca}
  \begin{subtable}[h]{\textwidth}
    \centering
    \caption{Main components identified.}
    \setlength{\tabcolsep}{0pt} 
    \begin{tabular*}{\columnwidth}{@{\extracolsep{\fill}\quad}lrrr@{}}
      \toprule
      \textbf{Factor} & \textbf{Comp. 1} & \textbf{Comp. 2} & \textbf{Comp. 3} \\
      \midrule
      BFI Conscientiousness                  & \textcolor{lightgray}{\( -0.022 \)} &                         \( 0.668 \) & \textcolor{lightgray}{\( -0.481 \)} \\
      BFI Neuroticism                        &                         \( 0.600 \) &                        \( -0.678 \) & \textcolor{lightgray}{\( -0.118 \)} \\
      ITQ Focus                              &                         \( 0.678 \) &  \textcolor{lightgray}{\( 0.203 \)} &                         \( 0.504 \) \\
      ITQ Involvement                        &                         \( 0.573 \) &                         \( 0.540 \) &  \textcolor{lightgray}{\( 0.417 \)} \\
      Exec. Time (delay 3.0 s, length 12)    &                         \( 0.758 \) & \textcolor{lightgray}{\( -0.178 \)} & \textcolor{lightgray}{\( -0.348 \)} \\
      Log EEG power \( \alpha + \beta \) Slice 1 & \textcolor{lightgray}{\( -0.436 \)} & \textcolor{lightgray}{\( -0.251 \)} &                         \( 0.589 \) \\
      \bottomrule
    \end{tabular*}
  \end{subtable}
  \newline
  \medskip
  \newline
  \begin{subtable}[h]{\textwidth}
    \centering
    \caption{Percentages of variance explained by the components.}
    \begin{tabular*}{\columnwidth}{@{\extracolsep{\fill}\quad}lrrrr@{}}
      \toprule
      {} & \textbf{Comp. 1} & \textbf{Comp. 2} & \textbf{Comp. 3} & \textbf{Total} \\
      \midrule
      Explained Variance & \SI{31.88}{\percent} & \SI{22.22}{\percent} & \SI{19.03}{\percent} & \SI{73.13}{\percent} \\
      \bottomrule
    \end{tabular*}
  \end{subtable}
\end{table}

The PCA produced three components that accounted for \SI{73.13}{\percent} of the variance in the six factors considered.
Component 1 included neuroticism, both ITQ scores, and execution time, indicating that more sensitive and immersed individuals tended to slow their responses under extended delay.
Component 2 included conscientiousness and the absence of neuroticism along with immersive involvement, indicating that efficient and secure participants tended to be more involved.
The third component had positive loadings on EEG in Slice 1 (first four steps of a block) and the focus component of the ITQ.\@
An additional analysis including heart rate added a component but improved the PCA fit by only \SI{4.5}{\percent}, indicating that any variability in this measure across individuals is unrelated to personality, immersiveness, or outcome.
A further analysis including GSR produced a solution in which GSR loaded with the first component, along with execution time.

\begin{figure}[h]
    \centering
    \includegraphics[width=.8\textwidth]{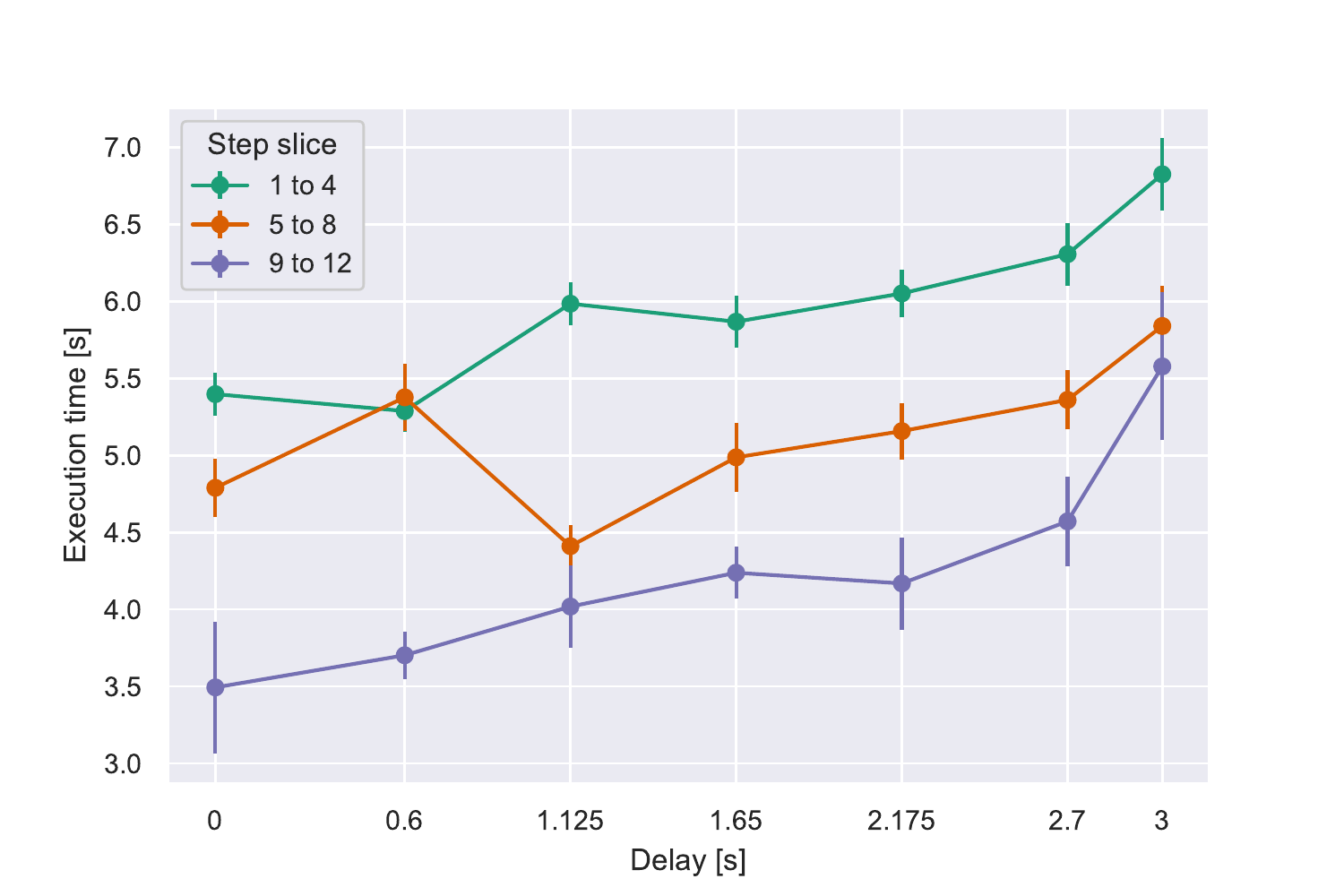}
    \caption{Correlation between neuroticism score of participants and their execution time in the longest block at the longest delay.
    Pearson correlation coefficient \( r = .40 \); 2-tailed \( p = .01 \).}
    \label{fig:neuro:exectime:reg}
\end{figure}

\cref{fig:neuro:exectime:reg} shows the correlation between neuroticism and execution time for the block with extreme values of length (12 steps) and delay (\SI{3.0}{\second}), for the full set of 40 participants.
It confirms the relationship indicated by the first component in the PCA using the smaller sample of 28 participants; that is, higher neuroticism is associated with more responsiveness to delay.

On the whole, these results suggest that individual differences in widely accepted personality variables and immersive tendencies moderate the response to delay.
This fact could have practical implications in the future.
It could, for instance, provide a tunable parameter for eventual models aiming to emulate human interaction with a WCA.\@
In addition, physiological measures of heart rate and EEG appear not to be direct indicators of  behavioral response to delay, although GSR may be more promising in this regard.

\section{Discussion}\label{sec:discussion}

We start our discussion with the main results of our experimentation presented in the previous section:

\begin{itemize}
  \item Firstly, and perhaps most importantly, we find that a system slow-down induces an \textit{additional} behavioral slow-down.
  That is, as system responsiveness decreases, our data indicates that users significantly slow-down in their execution of the task.
  This slow-down scales with the decrease in responsiveness; compared to the no-delay case, participants were on average \SI{12}{\percent} slower at \SI{1.65}{\second} delay and \SI{26}{\percent} at \SI{3.0}{\second} delay.
  Moreover, there is a temporal component to this effect; users become progressively slower the more time passes with reduced system responsiveness.

  \item Secondly, we find that the effects of behavioral slow-down due to impaired system responsiveness \emph{remain} for at least a few steps after system responsiveness improves.
  This is evidenced by the longer per-step-execution times of the first four steps of blocks immediately following a high-delay block, as pictured in \cref{fig:exectime:transition}.
  The question of whether any lingering effect can be measured after these four steps remains open.

  \item Thirdly, we evidence a speed-up in execution time over a series of steps; that is, subjects get faster at performing steps as the task progresses.
  However, the strength of this effect decreases as delay increases.
  Whereas for blocks without delay users performed the last four steps of a 12-step block on average \SI{36}{\percent} faster than the first four, at the maximum delay this effect practically disappears.

  \item Fourthly, in terms of inter-subject differences, PCA revealed three main factors governing users' response to delay. 
  The first factor represents sensitivity to delay as moderated by the ``Big Five'' personality trait of neuroticism and both measures of immersion: focus and involvement.
  Factor two and three represent dedication to the task as opposed to delay intolerance and reflect variables related to attentiveness, respectively.
  In simple terms, these results suggest that the effects of delay are most potent in individuals who are sensitive and involved in the task.
  The findings appear selective to cognitive assistance tasks like the present ones, inasmuch as the same measures did not correlate with outcomes in other computer-intensive environments such as immersive VR~\autocite{quesnel2018you}.
  These correlations are also consistent with previous findings indicating that individuals scoring high in neuroticism tend to be intolerant to delayed reward~\autocite{hirsh2008delay}.
\end{itemize}

A central question therefore arises: to which physio- and psychological mechanisms can these findings, most importantly the substantial slow-down in task execution, be attributed?

In \cref{sec:background}, we initially considered the possibility that delays might produce negative emotional reactions.
These could in turn elicit generalized arousal.
We also postulated that adapting to delay might progressively deplete cognitive resources in users.
However, the present data provide relatively little support for these alternatives, in that physiological measures of GSR and HR failed to show evidence of differential arousal under long vs.\ short delays, and speed-induced errors and non-completions predicted by resource depletion were not observed.
The acceleration data further do not indicate that extended delay increases erratic movement.
To the contrary, the data suggest this effect results from a delay in movement after an instruction is introduced.
That is, users fail to capitalize on the new information as quickly as they could.
Thus, contrary to our preliminary postulations, behavioral effects seem to arise from impaired cognitive control mechanisms, and not from emotion or resource depletion.
We hypothesize that the effects of feedback latency can best be understood as changes in the use of a cognitive plan.
As was described in \cref{sec:experimentaldesign,fig:lego:hierarchical}, complex cognitive and motor tasks have been modeled as the unfolding of a hierarchy of command, from high-level plans to physical output.
Long system latencies, we propose, disrupt the automating of such a plan, instead relegating it to attention-based control at the step-by-step level that is easily diverted.
This also provides a possible explanation for the lingering effects of delay after an acceptable system responsiveness is restored, as users needs time re-adjust and re-automate their cognitive plans.

As to the applicability of our findings to other applications, it must be noted that these results pertain to a specific class of applications, namely step-based task-guidance WCA.\@
However, we would expect our findings to extend to similar applications, as long as they follow the same pattern of seamless interaction --- i.e.\ such that the user does not need explicitly interact with the application to advance the state.

The results here presented provide a number of possible implications for WCA system design and optimization, both for single and multi-application flows.

\begin{itemize}

  \item Due to the behavioral slow-down in users, even short-term reductions in responsiveness will lead to significantly extended application lifetimes.
  This has direct implications for resource and power consumption.
  
  \item The fact that the adverse effects of delay on users do not immediately disappear as the system returns to a high-responsive state could have unconventional consequences for resource allocation.
  This is of particular importance, for instance, for cases where the user may be able to finish the task before these effects subside.
  In such cases, the limited potential gains might not justify diverting valuable system resources to the impaired application.

  \item In multi-user environments, the time dependency of user slow-down effects mean that fair degradation of system responsiveness across applications may not ultimately be beneficial to the system as a whole.
  Take for instance two applications on the same system which negatively interfere with each other. 
  The longer they interfere with each other, the longer their respective lifetimes are going to be, which in turn causes them to interfere even longer, potentially entering a negative feedback loop. 
  In such a case, prioritizing one over the other rather than trying to improve responsiveness for both might lead to resources being freed up faster system-wide.

  \item Based on our findings relating individual differences between users and their sensitivity to delays, it might also be possible to extrapolate user characteristic from measured execution times.
  This could prove a valuable tool for load balancing, for instance by prioritizing resource allocation to users with a higher sensitivity to system-state degradation.
  However, this remains an open challenge.

\end{itemize}

To wrap up, we believe the present data provide novel and unexpected insights for the understanding and optimizing of WCA deployments.
Although more subtle than expected, and in some cases somewhat counterintuitive, these insights represent a valuable tool to tackle inefficiencies in these systems.
Moreover, we also argue these findings represent a first step towards a full-fledged understanding of the relationship between application responsiveness and human behavior.
More research in this area will surely uncover more complex and interesting behaviors.
Finally, we believe the data provide parameters that can usefully be integrated into cognitive models of WCA that might be constructed under existing architectures like ACT-R.
These same parameters could be used to modulate the timing and generation of inputs in trace-based workload generation tools such as the EdgeDroid platform~\autocite{olguin:2018,olguin:2019}, allowing the tool to use the same trace to generate workloads for a multitude of different user profiles.

\section{Conclusion}\label{sec:conclusion}

In this paper, we presented the results of a study on the physiological and behavioral reactions of users of WCA to delays in the application pipeline.
Our ultimate aim was to identify and categorize the ways in which humans react to low system responsiveness in step-based cognitive assistance systems.

We approached this in an experimental manner, by having participants interact with an instrumented WCA setup, and found that delay appears to affect the cognitive plan of users, preventing them from automating the task they are performing.
The results show that user interactions in a WCA slow down in the presence of delays in the application pipeline.
When system responsiveness is high, the user responds quickly; when it is low, the user slows down.
This was evidenced by an increase in task execution times, even after accounting for the artificially introduced delays, as well as accelerometer data from participants wrists.
Additionally, we found that the strength of this effect is modulated by individual differences between subjects.

These results are interesting as they open up hitherto unexplored opportunities for the design, optimization and benchmarking of WCA systems.
In this context, we believe there are two direct and important next steps to be performed.
The first of these relates to the implications for system optimization and resource allocation discussed in \cref{sec:discussion}.
We believe these need to be implemented, tested and validated in real setups.
In particular, we identify two of these implications to be prime candidates for their own experimental studies.
One, our postulation that in cases where an impaired application is close to finishing, diverting resources to it might not be the most optimal course of action; and two, the possibility that ``fair'' degradation of system responsiveness across applications may be, in some cases, undesirable.
Both of these questions could be answered with straightforward setups.

The second step corresponds to the extension of existing tools for WCA benchmarking with the findings presented in this paper.
These tools are of great usefulness for the study of WCA systems, as they allow for automated large-scale testing without having to resort to human users.
However, they are still somewhat simplistic and unrealistic in their workload generation schemes.
Incorporating the results presented here would allow for much more realistic workloads.
For instance, our findings relating to the effects of delay on execution times could be directly adapted to modulate timings in the input stream.
Another example would be employing the results linking neuroticism to a heightened sensibility for delays to provide a ``tuning knob'' for the user models in these workload generation schemes.
Extending and perfecting these tools will allow for much more realistic benchmarking and testing of WCA systems, providing data of significantly better quality and ultimately leading to faster improvement, optimization and adoption of these systems.

\begin{acks}
  This research was supported in part by the Swedish Foundation for Strategic Research (SSF) under grant \texttt{ITM17-0246} (ExPECA) and the United States National Science Foundation (NSF) under grant number \texttt{CNS-1518865}.
  
  Any opinions, findings, conclusions or recommendations expressed in this material are those of the authors and do not necessarily reflect the view(s) of their employers or the above-mentioned funding sources.
\end{acks}


%
%
%

\printbibliography%

@String{Computing = "Computing" }

@String{Computer = "{IEEE} Computer" }

@String{Macmillan = "Macmillan" }

@String{Springer = "Springer-Verlag" }

@inproceedings{Ha:TowardsWearableCogAssist,
    author = {K. Ha and Z. Chen and W. Hu and W. Richter and P. Pillai and M. Satyanarayanan},
    title = {Towards Wearable Cognitive Assistance},
    booktitle = {Proceedings of the 12th Annual International Conference on Mobile Systems, Applications, and Services},
    series = {MobiSys '14},
    year = {2014},
    isbn = {978-1-4503-2793-0},
    location = {Bretton Woods, New Hampshire, USA},
    pages = {68--81},
    numpages = {14},
    url = {http://doi.acm.org/10.1145/2594368.2594383},
    doi = {10.1145/2594368.2594383},
    acmid = {2594383},
    publisher = {ACM},
    address = {New York, NY, USA},
}

@inproceedings{Chen:EarlyImplementation,
    author = {Z. Chen and L Jiang and W. Hu and K. Ha and B. Amos and P. Pillai and A. Hauptmann and M. Satyanarayanan},
    title = {Early Implementation Experience with Wearable Cognitive Assistance Applications},
    booktitle = {Proceedings of the 2015 Workshop on Wearable Systems and Applications},
    series = {WearSys '15},
    year = {2015},
    isbn = {978-1-4503-3500-3},
    location = {Florence, Italy},
    pages = {33--38},
    numpages = {6},
    url = {http://doi.acm.org/10.1145/2753509.2753517},
    doi = {10.1145/2753509.2753517},
    acmid = {2753517},
    publisher = {ACM},
    address = {New York, NY, USA},
}

@inproceedings{Chen:AnEmpiricalStudyOfLatency,
    author = {Z. Chen and W. Hu and J. Wang and S. Zhao and B. Amos and G. Wu and K. Ha and K. Elgazzar and P. Pillai and R. Klatzky and D. Siewiorek and M. Satyanarayanan},
    title = {An Empirical Study of Latency in an Emerging Class of Edge Computing Applications for Wearable Cognitive Assistance},
    booktitle = {Proceedings of the Second ACM/IEEE Symposium on Edge Computing},
    series = {SEC '17},
    year = {2017},
    isbn = {978-1-4503-5087-7},
    location = {San Jose, California},
    pages = {14:1--14:14},
    articleno = {14},
    numpages = {14},
    url = {http://doi.acm.org/10.1145/3132211.3134458},
    doi = {10.1145/3132211.3134458},
    acmid = {3134458},
    publisher = {ACM},
    address = {New York, NY, USA},
}

@online{openbci:headbandkit,
    title = {{OpenBCI EEG Headband Kit}},
    journal = {{Open Source Tools for Neuroscience}},
    year = {2020},
    month = {Mar},
    note = {[Online; accessed 20 Mar 2020]},
    url = {https://shop.openbci.com/products/openbci-eeg-headband-kit?variant=8120393760782},
}

@online{openbci:ganglion,
    title = {{OpenBCI Ganglion Biosensing Board}},
    journal = {{Open Source Tools for Neuroscience}},
    year = {2020},
    month = {Mar},
    note = {[Online; accessed 20 Mar 2020]},
    url = {https://shop.openbci.com/products/ganglion-board?variant=13461804483},
}

@online{empatica:e4,
    title = {{Empatica E4}},
    journal = {{Empatica}},
    year = {2020},
    month = {Mar},
    note = {[Online; accessed 24 Mar 2020]},
    url = {https://www.empatica.com/research/e4/},
}

@INPROCEEDINGS{olguin:2018,
  author={M. {Olguín Muñoz} and J. {Wang} and M. {Satyanarayanan} and J. {Gross}},
  booktitle={2018 IEEE/ACM Symposium on Edge Computing (SEC)}, 
  title={Demo: Scaling on the Edge --- A Benchmarking Suite for Human-in-the-Loop Applicationss}, 
  year={2018},
  volume={},
  number={},
  pages={323-325}
}

@inproceedings{olguin:2019,
    author = {M. {Olguín Muñoz} and J. Wang and M. Satyanarayanan and J. Gross},
    title = {EdgeDroid: An Experimental Approach to Benchmarking Human-in-the-Loop Applications},
    year = {2019},
    isbn = {9781450362733},
    publisher = {Association for Computing Machinery},
    address = {New York, NY, USA},
    url = {https://doi.org/10.1145/3301293.3302353},
    doi = {10.1145/3301293.3302353},
    booktitle = {Proceedings of the 20th International Workshop on Mobile Computing Systems and Applications},
    pages = {93–98},
    numpages = {6},
    location = {Santa Cruz, CA, USA},
    series = {HotMobile '19}
}

@article{dabrowsky:2011:40years,
    author = {J. Dabrowski and E. V. Munson},
    title = {{40 years of Searching for the Best Computer System Response Time}},
    year = {2011},
    issue_date = {September 2011},
    publisher = {Elsevier Science Inc.},
    address = {USA},
    volume = {23},
    number = {5},
    issn = {0953-5438},
    url = {https://doi.org/10.1016/j.intcom.2011.05.008},
    doi = {10.1016/j.intcom.2011.05.008},
    journal = {Interact. Comput.},
    month = sep,
    pages = {555–564},
    numpages = {10}
}

@article{eeg1020system:1961,
    title = {The Ten Twenty Electrode System: International Federation of Societies for Electroencephalography and Clinical Neurophysiology},
    journal = {American Journal of EEG Technology},
    volume = {1},
    number = {1},
    pages = {13-19},
    year  = {1961},
    publisher = {Taylor & Francis},
    doi = {10.1080/00029238.1961.11080571},
    URL = {https://doi.org/10.1080/00029238.1961.11080571},
    eprint = {https://doi.org/10.1080/00029238.1961.11080571}
}

@article{witmer1998:itq,
  title={Measuring presence in virtual environments: A presence questionnaire},
  author={B. G. Witmer and M. J. Singer},
  journal={Presence},
  volume={7},
  number={3},
  pages={225--240},
  year={1998},
  publisher={MIT Press}
}

@article{john1999:bfi,
  title={The Big Five trait taxonomy: History, measurement, and theoretical perspectives},
  author={O. P. John and S. Srivastava and others},
  journal={Handbook of personality: Theory and research},
  volume={2},
  number={1999},
  pages={102--138},
  year={1999},
  publisher={Guilford}
}

@inproceedings{khawadi2015:usinggsrtrust,
    author = {A. Khawaji and J. Zhou and F. Chen and N. Marcus},
    title = {{Using Galvanic Skin Response (GSR) to Measure Trust and Cognitive Load in the Text-Chat Environment}},
    year = {2015},
    isbn = {9781450331463},
    publisher = {Association for Computing Machinery},
    address = {New York, NY, USA},
    url = {https://doi-org.focus.lib.kth.se/10.1145/2702613.2732766},
    doi = {10.1145/2702613.2732766},
    booktitle = {Proceedings of the 33rd Annual ACM Conference Extended Abstracts on Human Factors in Computing Systems},
    pages = {1989–1994},
    numpages = {6},
    location = {Seoul, Republic of Korea},
    series = {CHI EA ’15}
}

@inproceedings{kuikkaniemi2010:biofeedback,
    author = {K. Kuikkaniemi and T. Laitinen and M. Turpeinen and T. Saari and I. Kosunen and N. Ravaja},
    title = {The Influence of Implicit and Explicit Biofeedback in First-Person Shooter Games},
    year = {2010},
    isbn = {9781605589299},
    publisher = {Association for Computing Machinery},
    address = {New York, NY, USA},
    url = {https://doi-org.focus.lib.kth.se/10.1145/1753326.1753453},
    doi = {10.1145/1753326.1753453},
    booktitle = {Proceedings of the SIGCHI Conference on Human Factors in Computing Systems},
    pages = {859–868},
    numpages = {10},
    location = {Atlanta, Georgia, USA},
    series = {CHI ’10}
}

@inproceedings{solovey2014:classifyingdriverworkload,
    author = {E. T. Solovey and M. Zec and E. A. {Garcia Perez} and B. Reimer and B. Mehler},
    title = {Classifying Driver Workload Using Physiological and Driving Performance Data: Two Field Studies},
    year = {2014},
    isbn = {9781450324731},
    publisher = {Association for Computing Machinery},
    address = {New York, NY, USA},
    url = {https://doi-org.focus.lib.kth.se/10.1145/2556288.2557068},
    doi = {10.1145/2556288.2557068},
    booktitle = {Proceedings of the SIGCHI Conference on Human Factors in Computing Systems},
    pages = {4057–4066},
    numpages = {10},
    location = {Toronto, Ontario, Canada},
    series = {CHI ’14}
}

@article{Son2010,
  doi = {10.1007/s12239-010-0065-6},
  url = {https://doi.org/10.1007/s12239-010-0065-6},
  year = {2010},
  month = jul,
  publisher = {Springer Science and Business Media {LLC}},
  volume = {11},
  number = {4},
  pages = {533--539},
  author = {J. Son and B. Reimer and B. Mehler and A. E. Pohlmeyer and K. M. Godfrey and J. Orszulak and J. Long and M. H. Kim and Y. T. Lee and J. F. Coughlin},
  title = {Age and cross-cultural comparison of drivers' cognitive workload and performance in simulated urban driving},
  journal = {International Journal of Automotive Technology}
}

@article{Healey2005,
  doi = {10.1109/tits.2005.848368},
  url = {https://doi.org/10.1109/tits.2005.848368},
  year = {2005},
  month = jun,
  publisher = {Institute of Electrical and Electronics Engineers ({IEEE})},
  volume = {6},
  number = {2},
  pages = {156--166},
  author = {J. A. Healey and R. W. Picard},
  title = {Detecting Stress During Real-World Driving Tasks Using Physiological Sensors},
  journal = {{IEEE} Transactions on Intelligent Transportation Systems}
}

@article{peterson1907psycho,
  title={Psycho-physical investigations with the galvanometer and pneumograph in normal and insane individuals},
  author={F. Peterson and C. G. Jung},
  journal={Brain},
  volume={30},
  pages={153--218},
  year={1907},
  publisher={Macmillan}
}

@article{Antonenko2010,
  doi = {10.1007/s10648-010-9130-y},
  url = {https://doi.org/10.1007/s10648-010-9130-y},
  year = {2010},
  month = apr,
  publisher = {Springer Science and Business Media {LLC}},
  volume = {22},
  number = {4},
  pages = {425--438},
  author = {P. Antonenko and F. Paas and R. Grabner and T. van Gog},
  title = {Using Electroencephalography to Measure Cognitive Load},
  journal = {Educational Psychology Review}
}

@inproceedings{Grimes2008,
  doi = {10.1145/1357054.1357187},
  url = {https://doi.org/10.1145/1357054.1357187},
  year = {2008},
  publisher = {{ACM} Press},
  author = {D. Grimes and D. S. Tan and S. E. Hudson and P. Shenoy and R. P. N. Rao},
  title = {Feasibility and pragmatics of classifying working memory load with an electroencephalograph},
  booktitle = {Proceeding of the twenty-sixth annual {CHI} conference on Human factors in computing systems  - {CHI} {\textquotesingle}08}
}

@article{bruno2016multiple,
  title={Multiple channels of visual time perception},
  author={A. Bruno and G. M. Cicchini},
  journal={Current opinion in behavioral sciences},
  volume={8},
  pages={131--139},
  year={2016},
  publisher={Elsevier}
}

@article{droit2011emotion,
  title={Emotion and time perception: effects of film-induced mood},
  author={S. Droit-Volet and S. L. Fayolle and S. Gil},
  journal={Frontiers in integrative neuroscience},
  volume={5},
  pages={33},
  year={2011},
  publisher={Frontiers}
}

@article{heron2012duration,
  title={Duration channels mediate human time perception},
  author={J. Heron and C. Aaen-Stockdale and J. Hotchkiss and N. W. Roach and P. V. McGraw and D. Whitaker},
  journal={Proceedings of the Royal Society B: Biological Sciences},
  volume={279},
  number={1729},
  pages={690--698},
  year={2012},
  publisher={The Royal Society}
}

@article{hirsh2008delay,
  title={Delay discounting: Interactions between personality and cognitive ability},
  author={J. B. Hirsh and D. Morisano and J. B. Peterson},
  journal={Journal of research in personality},
  volume={42},
  number={6},
  pages={1646--1650},
  year={2008},
  publisher={Elsevier}
}

@article{matthews2011stimulus,
  title={Stimulus repetition and the perception of time: The effects of prior exposure on temporal discrimination, judgment, and production},
  author={W. J. Matthews},
  journal={PLoS one},
  volume={6},
  number={5},
  year={2011},
  publisher={Public Library of Science}
}

@book{nielsen1994usability,
  title={Usability engineering},
  author={J. Nielsen},
  year={1994},
  publisher={Morgan Kaufmann}
}

@book{seow2008designing,
  title={Designing and engineering time: The psychology of time perception in software},
  author={S. C. Seow},
  year={2008},
  publisher={Addison-Wesley Professional}
}

@book{shneiderman2016designing,
  title={Designing the user interface: strategies for effective human-computer interaction},
  author={B. Shneiderman and C. Plaisant and M. Cohen and S. Jacobs and N. Elmqvist and N. Diakopoulos},
  year={2016},
  publisher={Pearson}
}

@article{wiener2011multiple,
  title={Multiple mechanisms for temporal processing},
  author={M. Wiener and M. S. Matell and H. Coslett},
  journal={Frontiers in integrative neuroscience},
  volume={5},
  pages={31},
  year={2011},
  publisher={Frontiers}
}

@incollection{zakay1996role,
  title={The role of attention in time estimation processes},
  author={D. Zakay and R. A. Block},
  booktitle={Advances in psychology},
  volume={115},
  pages={143--164},
  year={1996},
  publisher={Elsevier}
}

@article{zakay1995attentional,
  title={An attentional-gate model of prospective time estimation},
  author={D. Zakay and R. A. Block},
  journal={Time and the dynamic control of behavior},
  pages={167--178},
  year={1995}
}

@inproceedings{ragot2017emotion,
  title={Emotion recognition using physiological signals: laboratory vs. wearable sensors},
  author={M. Ragot and N. Martin and S. Em and N. Pallamin and J. M. Diverrez},
  booktitle={International Conference on Applied Human Factors and Ergonomics},
  pages={15--22},
  year={2017},
  organization={Springer}
}

@inproceedings{mccarthy2016validation,
  title={Validation of the Empatica E4 wristband},
  author={C. McCarthy and N. Pradhan and C. Redpath and A. Adler},
  booktitle={2016 IEEE EMBS International Student Conference (ISC)},
  pages={1--4},
  year={2016},
  organization={IEEE}
}

@article{agarwal2017eeg,
  title={EEG signal enhancement using cascaded S-Golay filter},
  author={S. Agarwal and A. Rani and V. Singh and A. P. Mittal},
  journal={Biomedical Signal Processing and Control},
  volume={36},
  pages={194--204},
  year={2017},
  publisher={Elsevier}
}

@article{savitzky1964smoothing,
  title={Smoothing and differentiation of data by simplified least squares procedures.},
  author={A. Savitzky and M. J. E. Golay},
  journal={Analytical chemistry},
  volume={36},
  number={8},
  pages={1627--1639},
  year={1964},
  publisher={ACS Publications}
}

@inproceedings{haapalainen2010psycho,
  title={Psycho-physiological measures for assessing cognitive load},
  author={E. Haapalainen and S. Kim and J. F. Forlizzi and A. K. Dey},
  booktitle={Proceedings of the 12th ACM international conference on Ubiquitous computing},
  pages={301--310},
  year={2010}
}

@article{kumar2016measurement,
  title={Measurement of cognitive load in HCI systems using EEG power spectrum: an experimental study},
  author={N. Kumar and J. Kumar},
  journal={Procedia Computer Science},
  volume={84},
  pages={70--78},
  year={2016},
  publisher={Elsevier}
}

@article{quesnel2018you,
  title={Are you awed yet? How virtual reality gives us awe and goose bumps},
  author={D. Quesnel and B. E. Riecke},
  journal={Frontiers in psychology},
  volume={9},
  pages={2158},
  year={2018},
  publisher={Frontiers}
}

@article{koch2009neural,
  title={Neural networks engaged in milliseconds and seconds time processing: evidence from transcranial magnetic stimulation and patients with cortical or subcortical dysfunction},
  author={G. Koch and M. Oliveri and C. Caltagirone},
  journal={Philosophical Transactions of the Royal Society B: Biological Sciences},
  volume={364},
  number={1525},
  pages={1907--1918},
  year={2009},
  publisher={The Royal Society London}
}

@article{lewis2003distinct,
  title={Distinct systems for automatic and cognitively controlled time measurement: evidence from neuroimaging},
  author={P. A. Lewis and R. C. Miall},
  journal={Current opinion in neurobiology},
  volume={13},
  number={2},
  pages={250--255},
  year={2003},
  publisher={Elsevier}
}

@article{lee2019limiting,
  title={Limiting motor skill knowledge via incidental training protects against choking under pressure},
  author={T. G. Lee and D. E. Acuña and K. P. Kording and S. T. Grafton},
  journal={Psychonomic bulletin \& review},
  volume={26},
  number={1},
  pages={279--290},
  year={2019},
  publisher={Springer}
}

@article{puttemans2005changes,
  title={Changes in brain activation during the acquisition of a multifrequency bimanual coordination task: from the cognitive stage to advanced levels of automaticity},
  author={V. Puttemans and N. Wenderoth and S. P. Swinnen},
  journal={Journal of Neuroscience},
  volume={25},
  number={17},
  pages={4270--4278},
  year={2005},
  publisher={Soc Neuroscience}
}

@article{jeon2015degree,
  title={Degree of automaticity and the prefrontal cortex},
  author={H. Jeon and A. D. Friederici},
  journal={Trends in cognitive sciences},
  volume={19},
  number={5},
  pages={244--250},
  year={2015},
  publisher={Elsevier}
}

@article{baumeister74tice,
  title={Tice (1998). Ego Depletion: Is the Active Self a Limited Resource},
  author={R. E. Baumeister and E. Bratslavsky and M. Muraven},
  journal={Journal of personality and social psychology},
  volume={74},
  number={5},
  pages={1252}
}

@article{lin2020strong,
  title={Strong effort manipulations reduce response caution: A preregistered reinvention of the ego-depletion paradigm},
  author={H. Lin and B. Saunders and M. Friese and N. J. Evans and M. Inzlicht},
  journal={Psychological science},
  pages={0956797620904990},
  year={2020},
  publisher={SAGE Publications Sage CA: Los Angeles, CA}
}

@article{neves1981knowledge,
  title={Knowledge Compilation: Mechanisms for the},
  author={D. M. Neves and J. R. Anderson},
  journal={Cognitive skills and their acquisition},
  volume={6},
  pages={57},
  year={1981},
  publisher={Psychology Press}
}

@article{pijpers2003anxiety,
  title={Anxiety--performance relationships in climbing: a process-oriented approach},
  author={J. R. R. Pijpers and R. R. D. Oudejans and F. Holsheimer and F. C. Bakker},
  journal={Psychology of sport and exercise},
  volume={4},
  number={3},
  pages={283--304},
  year={2003},
  publisher={Elsevier}
}

@ARTICLE{funk2015caworkplace,
  author={M. {Funk} and A. {Schmidt}},
  journal={IEEE Pervasive Computing}, 
  title={Cognitive Assistance in the Workplace}, 
  year={2015},
  volume={14},
  number={3},
  pages={53-55}
}

@inproceedings{gorecky2011cognito,
  author = {D. Gorecky and S. F. Worgan and G. Meixner},
  title = {{COGNITO: A Cognitive Assistance and Training System for Manual Tasks in Industry}},
  year = {2011},
  isbn = {9781450310291},
  publisher = {Association for Computing Machinery},
  address = {New York, NY, USA},
  url = {https://doi.org/10.1145/2074712.2074723},
  doi = {10.1145/2074712.2074723},
  booktitle = {Proceedings of the 29th Annual European Conference on Cognitive Ergonomics},
  pages = {53-6},
  numpages = {4},
  location = {Rostock, Germany},
  series = {ECCE '11}
}

@article{Loewenstein1992anomaliesintertemporalchoice,
  author = {G. Loewenstein and D. Prelec},
  title = "{Anomalies in Intertemporal Choice: Evidence and an Interpretation*}",
  journal = {The Quarterly Journal of Economics},
  volume = {107},
  number = {2},
  pages = {573-597},
  year = {1992},
  month = {05},
  issn = {0033-5533},
  doi = {10.2307/2118482},
  url = {https://doi.org/10.2307/2118482},
  eprint = {https://academic.oup.com/qje/article-pdf/107/2/573/5203125/107-2-573.pdf},
  }

@online{IKEAAssistant,
    title = {{IKEA assistant}},
    journal = {Elijah Group at Carnegie Mellon},
    year = {2020},
    month = {Jul},
    note = {[Online; accessed 24. Jul. 2020]},
    url = {https://www.youtube.com/watch?v=vMzTGgQHjpo}
}

@online{PingPongAssistant,
  title = {{Ping Pong assistant}},
  journal = {Elijah Group at Carnegie Mellon},
  year = {2020},
  month = {Jul},
  note = {[Online; accessed 24. Jul. 2020]},
  url = {https://www.youtube.com/watch?v=_lp32sowyUA}
}
\end{document}